\documentclass[aps,pra,twocolumn,showpacs,superscriptaddress]{revtex4}
\usepackage{graphicx}
\usepackage{dcolumn}
\usepackage{color}
\usepackage{bm}
\usepackage{amssymb}
\usepackage{amsmath}

\usepackage{hyperref} 

\begin{document}

\title{Dependence of in-situ Bose condensate size on frequency of RF-field used for evaporative cooling}

\author{S. R. Mishra}
\email[E-mail: ]{srm@rrcat.gov.in}
\affiliation{Raja Ramanna Centre for Advanced Technology, Indore-452013, India.}
\affiliation{Homi Bhabha National Institute, Mumbai-400094, India.}
\author{S. P. Ram}
\author{S. K. Tiwari}
\affiliation{Raja Ramanna Centre for Advanced Technology, Indore-452013, India.}
\author{H. S. Rawat}
\affiliation{Raja Ramanna Centre for Advanced Technology, Indore-452013, India.}
\affiliation{Homi Bhabha National Institute, Mumbai-400094, India.}

\begin{abstract}

We report the results of in-situ characterization of $ ^{87}$Rb atom cloud in a quadrupole Ioffe configuration (QUIC) magnetic trap after radio frequency (RF) evaporative cooling of the trapped atom cloud. The in-situ absorption images of the atom cloud have shown clear bimodal optical density (OD) profiles which indicate the Bose-Einstein condensation (BEC) phase transition in the trapped gas. Also, we report the measured variation in the sizes of the condensate and thermal clouds with the final frequency in the frequency scan of the RF-field applied for evaporative cooling. The results are consistent with the theoretical understanding and predictions reported earlier.

\end{abstract}


 
\maketitle


\section{Introduction}
Over more than two decades, the laser atom cooling \cite{Adams1997a} and Bose-Einstein condensation (BEC) of dilute atomic gases  \cite{Anderson1995a, Inguscio1999a} are under intense investigation for interesting physics as well as for various applications. With the advent of techniques of laser cooling and trapping of atoms, such as magneto-optical trap (MOT), an unprecedented control on number and temperature in an atomic sample has been possible, which has proved useful in various atomic physics experiments \cite{Anderson1995a, Inguscio1999a} and technological applications such as atom-lithography \cite{Dwyer2005a}, accurate atomic clocks \cite{Ludlow2015}, cold-atom gyroscopes \cite{Gauguet2009a,FangQin2012}, cold-atom accelerometers and gravimeters \cite{Bodart2010gravimeter}, atomic magnetometers \cite{Behbood2013}, etc. Cold atoms trapped in a periodic potential formed by the standing wave pattern of far detuned laser beams, known as optical lattice, mimic the electrons in a periodic potential of a crystal lattice. Because of opportunity to tailor the potential of optical lattices in several ways, cold atoms and Bose condensate of atoms trapped in optical lattices serve as a test bed to model and verify various condensed matter phenomena with a greater flexibility than before \cite{Lewenstein2007a,Inguscio2013} and also provide opportunity to explore dynamics of matter-waves \cite{Morsch2006,Bloch2008a,Lewenstein2007a}. Further, cold atoms in optical lattices are also considered promising systems for future technology of quantum information processing \cite{Bloch2012a}. The cooling and trapping of atoms has also enriched our basic understanding of atomic physics which includes behaviour of fermions and bosons confined in different dimensions and geometries, Feshbach resonances \cite{Inouye1998a, Chin2010a}, dressed states of atom in strong optical \cite{Cohen-Tannoudji2011a} and radio-frequency \cite{Zobay2001,Morizot2006a,Chakraborty2016a} fields. 

Thus preparation of cold atoms and Bose condensate samples of atomic gases is the first step to proceed towards different uses as discussed above.
A magneto-optical trap (MOT) is a robust and reliable technique to produce cold atomic samples in the temperature range of 10-100 $\mu K$ with number density in the range of $10^{10}-10^{11}$ cm$^{-3}$. To obtain a Bose condensate of the atoms, for example in $^{87}$Rb atom cloud, a further lower temperature (in the sub-micro-Kelvin range) and higher number density (in the range $10^{13}-10^{14}$ cm$^{-3}$) is needed. This comes from the requirement of the phase-space density ( $\rho = n\lambda_{dB}^{3}$ ) of the atom cloud to be greater than 2.61 for BEC, where $n$ is number density of atoms and $ \lambda_{dB} $ is de-Broglie wavelength. Due to dissipative processes in resonant interaction of atoms with light, it is difficult to achieve such a high value of phase space-density in the atom cloud in a MOT. Thus to achieve Bose-Einstein condensation (BEC) in a dilute atomic gas, the laser cooling in a MOT is used as the first stage of cooling which is succeeded by the second stage cooling known as evaporative cooling \cite{Hess1986a,Ketterle1996181, Anderson1996a,Inguscio1999a}. There are several variants of design of an experimental setup to implement the above two-stage cooling protocol for achieving BEC. The second stage cooling, i.e. evaporative cooling, is performed while atoms are trapped in a conservative potential of either a magnetic trap or a dipole trap of a far detuned laser beam \cite{Barrett2001a}. For evaporative cooling, an ultrahigh vacuum (UHV) environment is necessary to have a long lifetime of atoms in the trap during the evaporative cooling. This UHV requirement conflicts with the requirement of background vapor needed for the MOT loading. One way to resolve this issue is to use the concept of double-MOT setup \cite{Myatt1996a, SRMishra2008a}, which implements the formation the first MOT in a vapor chamber and the second MOT in an UHV chamber by transferring the cold atoms from the first MOT (i.e. vapor chamber MOT). The vapor chamber and UHV chamber are connected, but differentially pumped to maintain the different levels of pressure. In the double-MOT setup, the MOT in the vapor chamber is called ``VC-MOT'', whereas the MOT in the UHV chamber is called ``UHV-MOT''. We have developed an experimental setup based on this double-MOT scheme to achieve the Bose-Einstein Condensation (BEC) of $^{87}$Rb atoms.

In this article, we report the in-situ characterization of RF-field induced evaporatively cooled atom cloud of $^{87}$Rb atoms in a Quadrupole Ioffe Configuration (QUIC) magnetic trap. The in-situ absorption images of the cooled atom cloud have shown clear bimodal optical density (OD) profiles which indicate the Bose-Einstein condensation (BEC) phase transition in the trapped gas. We also report, for the first time to the best of our knowledge, the measured variation in the sizes of the condensate and thermal clouds with the final frequency in the frequency scan of the RF-field applied for evaporative cooling. The results are consistent with the theory. These in-situ results can be useful to characterize the Bose condensate when it is difficult to switch-off the trap for the time-of-flight observations. 
 
This article is organized as follows. In section \ref{Experimental_setup}, we discuss our double-MOT setup developed for the realization of BEC of $^{87}$Rb atoms in our lab. In this section, vacuum chambers for formation of both the MOTs as well as for magnetic trapping, lasers and optical layout, coils used for MOT and magnetic traps and controller system for the setup are discussed. In section \ref{expt_procedure}, the procedure of formation of two MOTs, preparation of UHV-MOT cloud for magnetic trapping, magnetic trapping, evaporative cooling and characterization of cooled atom cloud by absorption probe imaging are discussed. Our main observations and results using the in-situ absorption imaging technique are presented in section \ref{results_discussion}. The results show the evidence of BEC phase transition in the evaporatively cooled cloud, and sizes of the condensate and the thermal clouds are found to vary with the final frequency of RF-field. Finally, we present the conclusions of this work in section \ref{conclusion}.

\begin{figure}[t]
	\centering
		\includegraphics[width = \columnwidth]{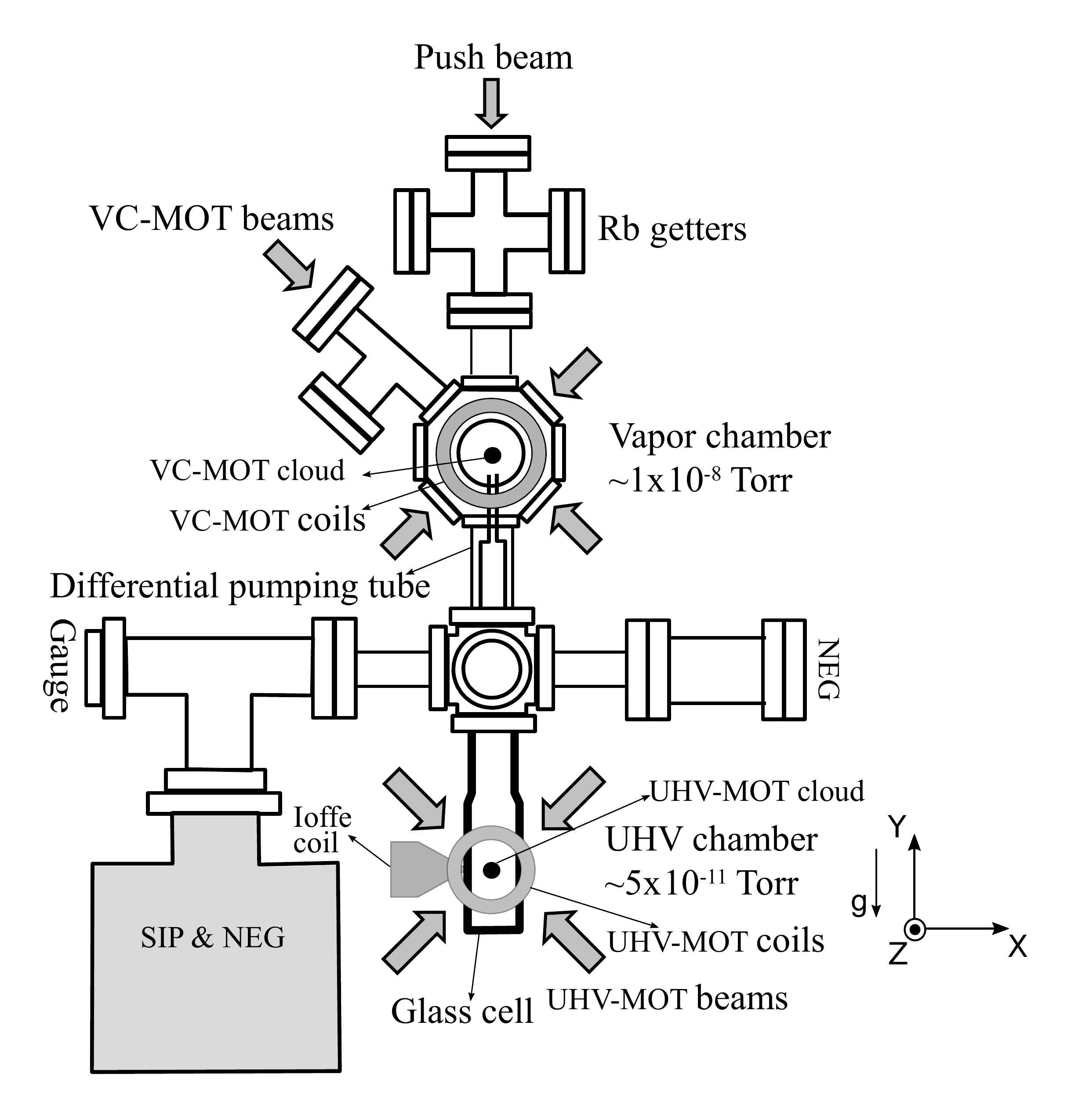}
	\caption{Schematic of the experimental setup. SIP: sputter-ion pump, NEG: non-evaporable getter pump.}
	\label{fig:Fig1Expt_setup}
\end{figure}

\section{Experimental Setup}
\label{Experimental_setup}

(a) Vacuum system:

The schematic of our experimental setup for Bose-Einstein condensation of $^{87}$Rb atoms is shown in Fig. \ref{fig:Fig1Expt_setup}. In this setup, an octagonal chamber of stainless-steel (SS) having Rubidium (Rb) vapor at $\sim 2\times10^{-8}$ Torr pressure (pressure without vapor $\sim 1\times10^{-8}$ Torr) is used for formation of vapor chamber MOT (VC-MOT). A quartz glass cell kept at a pressure of $\sim 5\times10^{-11}$ Torr, referred as UHV-MOT chamber, is used for loading UHV-MOT. A narrow tube of length 122 mm, called differential pumping tube (DPT), is connected between the cell and the octagonal chamber to ensure the differential pressure between them. The upper part (towards the VC-MOT chamber) of this tube has diameter of 2.5 mm upto 60 mm length and the remaining part has the diameter of 5 mm. This design is adapted to achieve the desired conductance as well as to accommodate the transverse expansion of the atom-flux during the transfer of atoms from VC-MOT to UHV-MOT. A six-way cross made from a cube of SS is used for connecting the octagonal VC-MOT chamber and the glass cell through the differential pumping tube, as shown in Fig.\ref{fig:Fig1Expt_setup}. The sides of this cube have appropriate knife-edges to connect other components with  NW35CF flanges. A combination sputter ion pump (VacIon plus, 150 l/s Starcell with a non-evaporable getter (NEG) module, from Varian, Italy) and a separate NEG pump (GP-100MK5, SORB-AC MK5 type cartridge pump from SAES getters, Italy) are also connected to the cube for pumping out the glass cell to the necessary UHV level. A turbo molecular pump (TMP) of capacity 70 l/s and a sputter ion pump of capacity 20 l/s are connected to the octagonal VC-MOT chamber. Bayerd-Alpert (BA) type ion gauges are used for measuring the pressure in the VC-MOT chamber and glass cell during the pumping. For measuring the very low pressure in the glass cell (UHV-MOT chamber), an extractor gauge (Oerlikon leybold), which can sense the pressure value upto $1\times 10^{-12}$ Torr, is connected to the chamber to which glass cell is connected. During the evacuation process, first of all the TMP is used for roughing and degassing of the whole vacuum system. A prolonged baking of the vacuum system for the duration of a week ($\sim$7x24 hours) at temperature of 100-150 $^o$C is performed for a good degassing of the whole system. During this baking, all the ion pumps and gauges are also baked and degassed. After cooling down of the vacuum system, all the ion pumps and NEG modules are activated and then TMP is switched-off and isolated from the vacuum system by using a gate-valve. With the ion pumps and NEG pumps ON, the pressure of ~$\sim1\times10^{-8}$ Torr in the octagonal chamber and ~$\sim5\times10^{-11}$ Torr in the glass cell are reached over a period of one to two days.

\begin{figure}[t]
	\centering
		\includegraphics[width = \columnwidth]{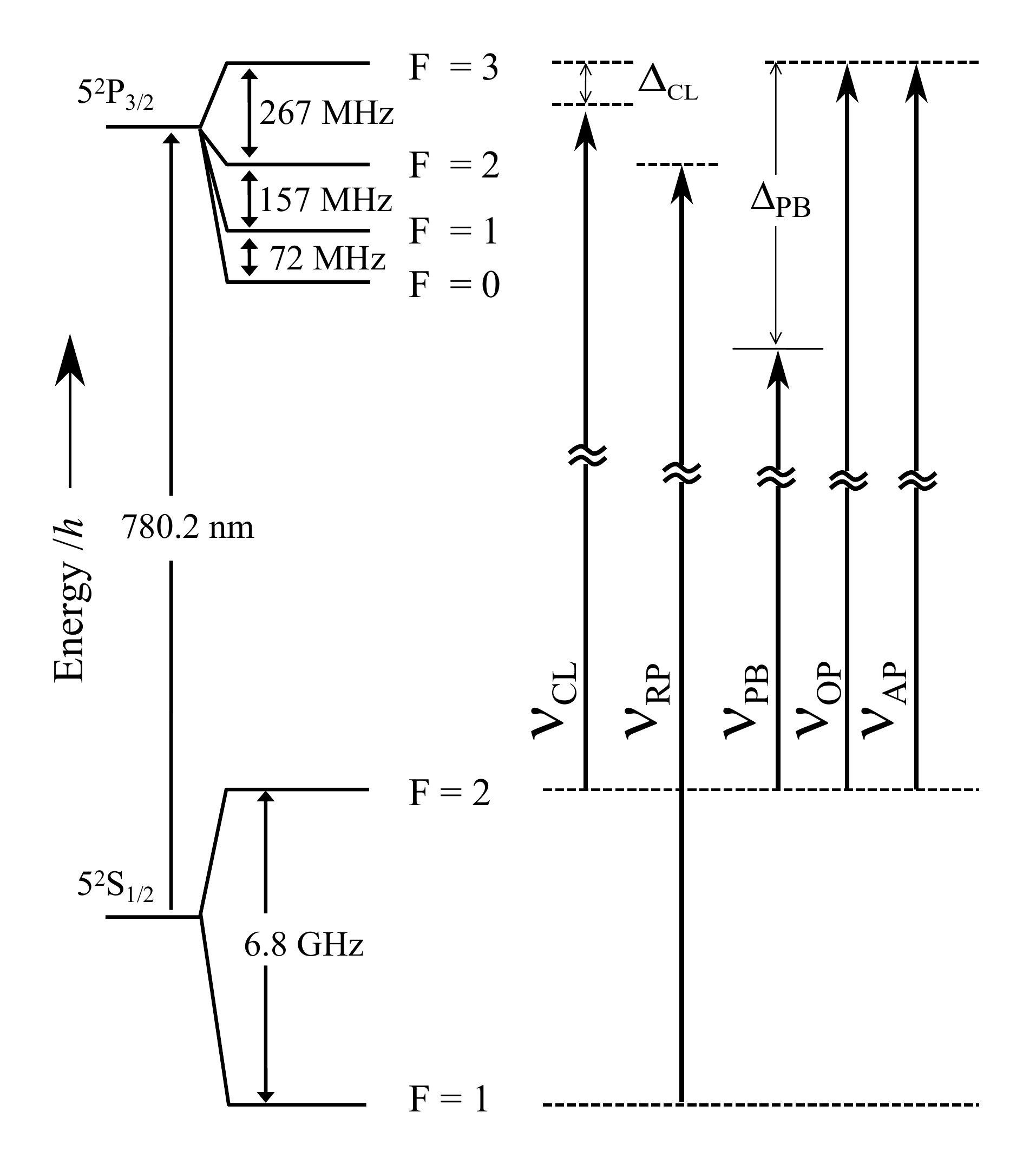}
	\caption{Schematic of the energy levels of $^{87}$Rb atom relevant to laser cooling and BEC experiments. The frequencies and corresponding detuning of various laser beams such as cooling ($\nu_{CL}$), repumping ($\nu_{RP}$), push ($\nu_{PB}$), optical pumping ($\nu_{OP}$), and absorption probe ($\nu_{AP}$) are indicated.}
	\label{fig:Fig2_rb_transition_level}
\end{figure}

(b) Laser systems and optical layout: 

\begin{figure}
     \centering
         \includegraphics[width = \columnwidth]{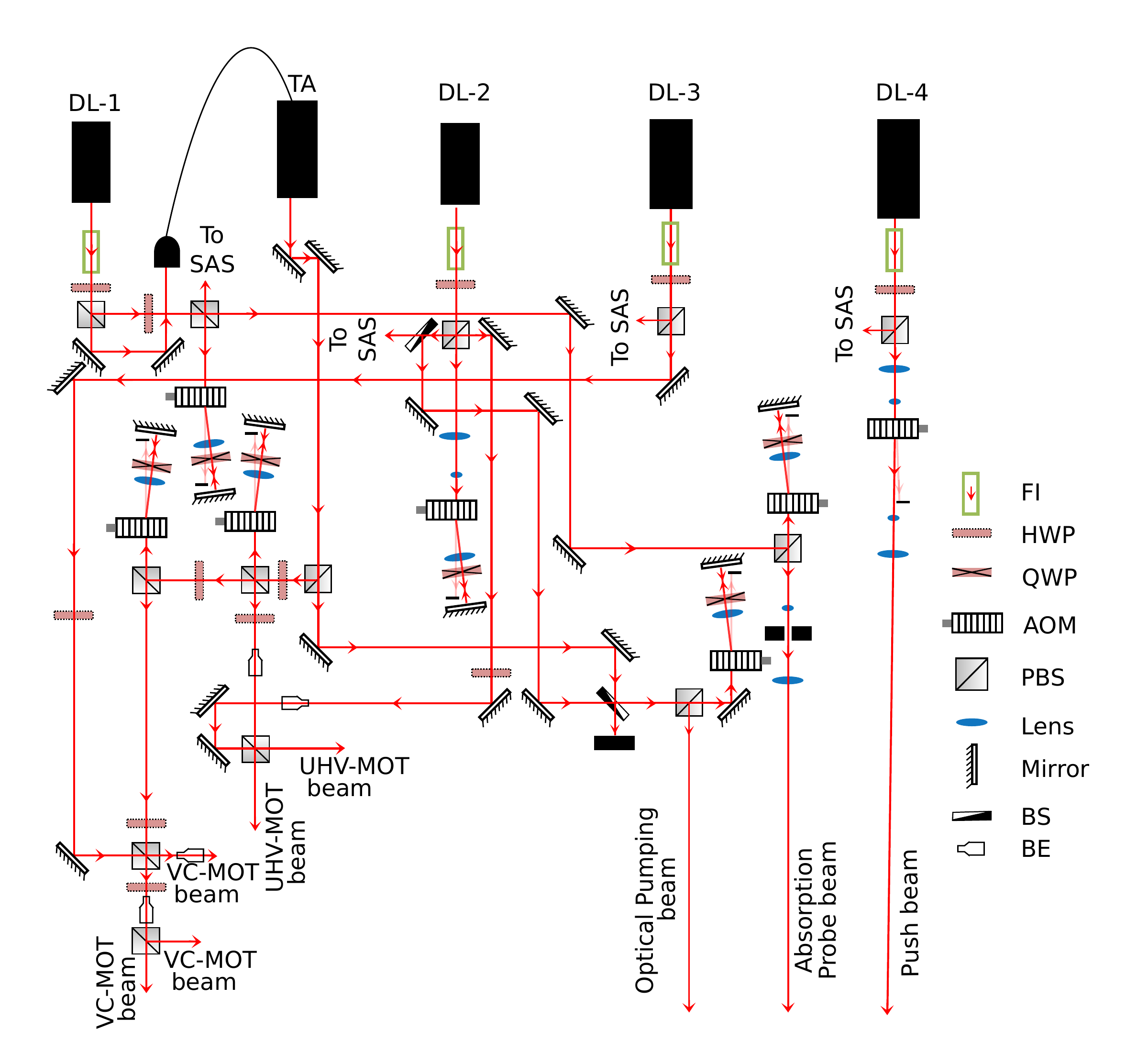}
     \caption{(Color online) Schematic of the optical layout for the generation of various laser beams required for MOT formation, optical pumping and detection. Various optical components used in the setup are abbreviated as FI: Faraday isolator, HWP: half-wave plate, QWP: quarter wave plate, AOM: acusto-optics modulator, PBS: polarizing cube beam splitter, BS: beam splitter and BE: beam expander.}
     \label{fig:Fig3_optical_layout}
\end{figure}

The relevant energy levels of $^{87}$Rb atom are as shown in Fig.\ref{fig:Fig2_rb_transition_level}. The cooling transition for $^{87}$Rb is F = 2 to F$^\prime$=3 and repumping transition is F=1 to F$^\prime$=2. The frequencies of various laser beams used in the experiments are set as shown schematically in Fig.\ref{fig:Fig2_rb_transition_level}. The lasers are frequency stabilized and locked using saturated absorption spectroscopy (SAS) technique. Different laser beams used for cooling, repumping, pushing, optical pumping and absorption imaging are derived from several external cavity diode lasers (ECDLs) operating at $\lambda \sim $780.2 nm. Their frequencies are referenced and locked according to the use. The cooling laser beams for VC-MOT and UHV-MOT are derived from the output beam of an oscillator-amplifier system. This system can deliver upto $\sim$1500 mW power in the output laser beam when a tapered amplifier (BOOSTA, Toptica, Germany) is seeded by an external cavity diode laser (ECDL) oscillator (DL-100, TOPTICA, Germany). The re-pumping laser beams for both the MOTs are obtained from two independent ECDL systems. The push beam is derived from another ECDL system. The absorption probe and optical pumping beams are derived from the cooling laser beams.

The Fig.\ref{fig:Fig3_optical_layout} shows the schematic of the optical layout of the components used for manipulation of the laser beams for various purposes. Various optical components such as polarizing beam splitters (PBSs), waveplates ($\lambda$/4 and $\lambda$/2), mirrors, lenses, acousto-optic modulators (AOMs), beam expanders (BEs), etc are used for different functions which include splitting, combining, setting the polarizations, reflecting, focusing, switching and controlling power, frequency shifting, expansion, etc. of the laser beams. The output from the BOOSTA amplifier (TA) is split to obtain the cooling beam for VC-MOT (power $\sim$ 250 mW), cooling beam for UHV-MOT (power $\sim$ 180 mW) and optical pumping beam (power $\sim$ 3 mW). The cooling beam for VC-MOT, after a double pass through an AOM, is further split and expanded to derive three VC-MOT beams (power $\sim$ 20 mW in each beam). The VC-MOT re-pumping laser beam from the laser DL-3 (power $\sim$ 15 mW) is mixed to one of three VC-MOT beams as shown in Fig.\ref{fig:Fig3_optical_layout}. These three VC-MOT beams are injected into the octagonal VC-MOT chamber in retro-reflection configuration to obtain the required six beam for VC-MOT formation. The quarter wave plates are used to set the appropriate polarization of the VC-MOT beams for MOT formation. The cooling beam for UHV-MOT, after a double pass through an AOM and expansion, is first mixed with a re-pumping laser beam and then split to obtain two UHV-MOT beams as Fig.\ref{fig:Fig3_optical_layout}. The repumping beam for UHV-MOT is derived from a 100 mW ECDL (DL-2) after double pass through an AOM as shown in Fig.\ref{fig:Fig3_optical_layout}. These two UHV-MOT beams are further split to obtain six independent UHV-MOT beams. All the six UHV-MOT beams have nearly equal power of $\sim$ 10 mW in the cooling part in each beam. The total power ( $\sim$ 20 mW) in repumping part, however, is not equally distributed in all the six UHV-MOT beams. This is because of different polarizations of cooling and re-pumping lasers after mixing and before splitting. Unlike the cooling power, the different repumping power in the six UHV-MOT beams does not cause any problem in the operation of UHV-MOT. However, six independent UHV-MOT beams with equal cooling power are suitable for stable UHV-MOT operation and for lowering the temperature of atom cloud during the molasses stage.

(c) Coils for formation of MOTs and magnetic trapping:
\begin{figure}[t]
	\centering
		\includegraphics[width = 0.7\columnwidth]{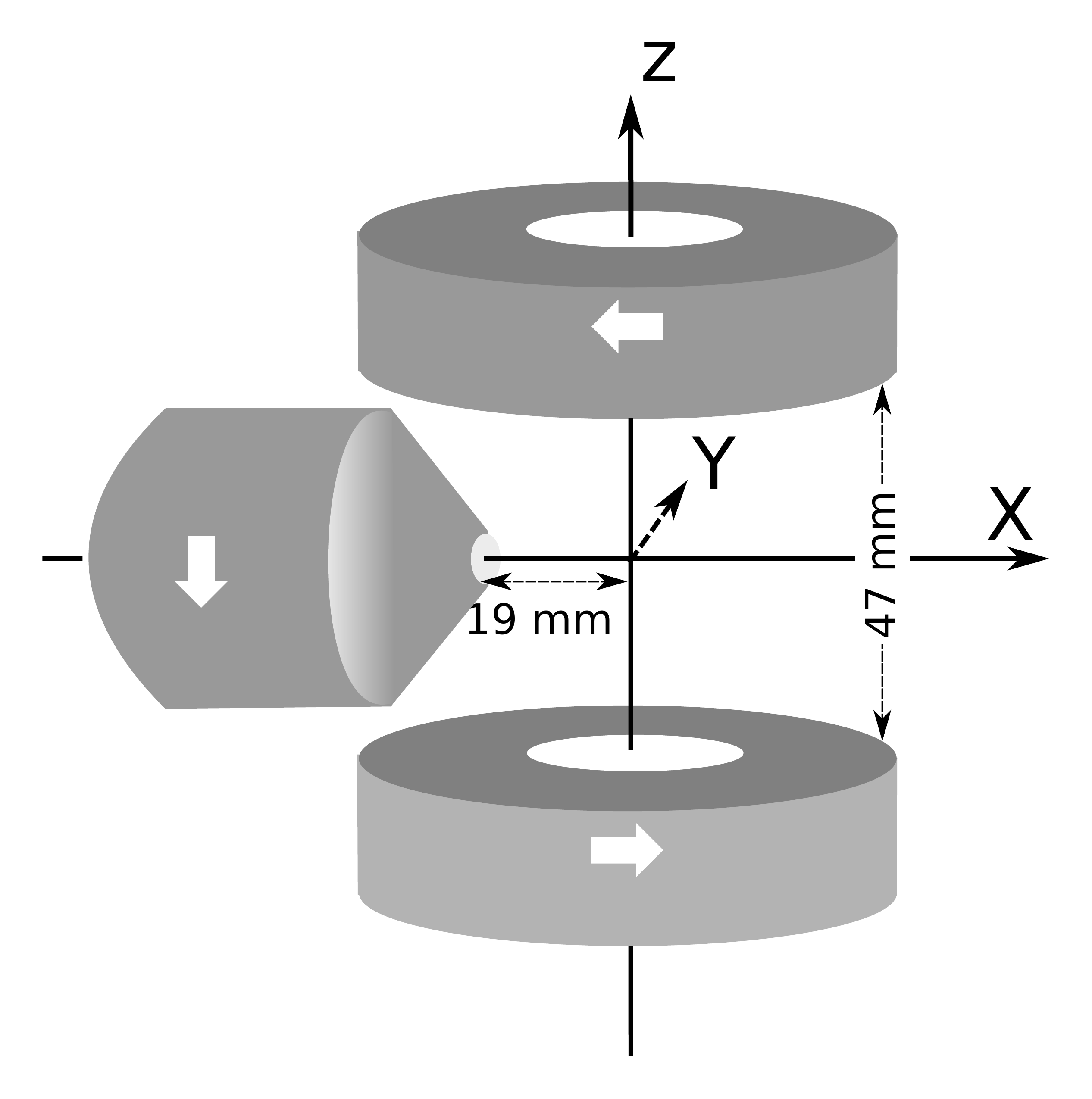}
	\caption{Schematic of configuration of QUIC trap coils. Separation between coils and the co-ordinate axes are shown. White arrows show the direction of the current in the coils.}
	\label{fig:QUIC_trap_coil}
\end{figure}

For the VC-MOT, a pair identical coils having currents in anti-Helmholtz configuration is used. Each coil has 150 number of turns of copper wire of diameter $\sim $ 1.5 mm, with size of coil as : inner diameter (ID) $\sim $ 57 mm, outer diameter (D) $\sim $ 66 mm and length $\sim $38 mm. The coils provide a quadrupole field gradient of $\sim $ 10 G/cm in the axial direction at DC current of 3 A in the coils for $\sim $ 55 mm face to face separation between the coils. For the UHV-MOT, another pair of coils, each coil having 230 number of turns (ID $\sim $ 34 mm, OD $\sim $ 70 mm, length $\sim $ 35 mm and wire diameter $\sim $ 1.4 mm), is used with face to face separation between coils $\sim $ 47 mm. With current in anti-Helmholtz configuration, these coils generate a quadrupole field with axial field gradient of $\sim$ 10 G/cm for a current of 1 A in each coil. These coils are kept inside suitably designed Teflon housing in which cold water is circulated for cooling of the coils. These UHV-MOT coils are also used as quadrupole trap coils for magnetic trapping of the atoms after initial cooling and trapping in the UHV-MOT. In the quadrupole magnetic trap, a high current ($>$20 A) flows in the coils, which makes the cooling of the coils necessary. The quadrupole magnetic trap is converted to a quadrupole-Ioffe configuration (QUIC) trap by including a third coil (called Ioffe coil) in the quadrupole trap setup. The Ioffe coil is placed at an axis passing through the quadrupole trap centre and perpendicular to the quadrupole trap axis, as shown in Fig. \ref{fig:QUIC_trap_coil}. The purpose of the QUIC trap is to have a magnetic trap with non-zero magnetic field at the minimum of the trap potential, which helps in reducing trap losses due to Majorana spin flips encounter in a quadrupole trap \cite{Esslinger1998a}.

(d) Controller system :

An industrial personal computer (PC) installed with a field programmable gate array (FPGA) card has been used to program and control the experimental procedure in the desired sequence. The durations and sequence of various events from VC-MOT formation to evaporative cooling and imaging are set using this controller system. The FPGA card is provided control through LabVIEW program. The generated signals from FPGA are appropriately amplified and used as trigger to various electronic devices such as power supplies, AOMs, mechanical shutters, CCD cameras, RF synthesizers and other circuitries. For the fast switching of the current in the magnetic trap coils, an Insulated Gate Bipolar Transistor(IGBT) based switching circuitry is used, which provides a fast (rise time $\sim $ 2.5 ms) switching of current in the coils. This circuitry also works after receiving an appropriate trigger pulse from the controller system. 

\section{Experimental procedure and measurements}
\label{expt_procedure}

In the experiments, first event is the formation of VC-MOT. The vapor of Rb atoms is generated in the octagonal VC-MOT chamber ( Fig. \ref{fig:Fig1Expt_setup}) by passing a dc current of 3-4 A through the Rb-getters inserted inside the chamber via a feed-through. A push laser beam is focussed on the VC-MOT atom cloud to eject the atoms from this MOT. The atoms ejected from the VC-MOT pass through the differential pumping tube (DPT) and reach the glass cell where they are recaptured in the UHV-MOT. The atoms in the UHV-MOT undergo the MOT compression, optical molasses and optical pumping stages, before trapping in the magnetic trap for evaporative cooling. A brief summary of sequence and duration of these processes and events during the experiments is presented as follows. 

(a) VC-MOT formation:

The Rb-vapor is generated in the octagonal VC-MOT chamber by passing a DC current of 2.7-3.5 A in the Rb-getter strips. The Rb-getter strips are fixed on a feed-through and inserted inside the chamber by mounting this feed-through on one of the ports of the octagonal VC-MOT chamber. The three VC-MOT beams, after making their cooling laser polarization circular using the quarter waveplates, are injected into the octagonal VC-MOT chamber through three viewports. These beams are retro-reflected by the mirrors kept at the other end of the chamber near the exit viewports. After retro-reflection of these three VC-MOT beams, six beams required for the MOT operation are achieved. Before retro-reflection from the mirror, each VC-MOT beam passes through a quarter waveplate to maintain the desired polarization of the beam after retro-reflection. A DC current of $\sim$ 3 A is passed through the VC-MOT coils to generate the required magnetic field gradient of $\sim$12 G/cm for MOT formation. The cooling laser frequency is locked at a side of the cooling transition (with red detuning of $\sim$2.5 $\Gamma$) whereas re-pumping laser is kept at the resonance of re-pumping transition of the $^{87}$Rb atom. After appropriate vapor pressure in the chamber ($\sim 2 \times10^{-8}$ Torr) and setting the laser beams frequency appropriately, we could trap $\sim 1\times10^8$ $^{87}$Rb atoms in the VC-MOT at temperature of $\sim$ 300 $\mu$K.

(b) Atom transfer and UHV-MOT formation:

The atoms cooled and trapped in the VC-MOT are utilized for loading the UHV-MOT in the glass cell. The VC-MOT and UHV-MOT are separated by a distance of $\sim$ 360 mm in our setup. The transfer of atoms from the VC-MOT to the UHV-MOT region, through the narrow differential pumping tube (DPT), is an important step in loading the UHV-MOT. It is always important to obtain the maximum number of atoms trapped in the UHV-MOT. This atom transfer from VC-MOT to UHV-MOT has been studied in detail by our group \cite{SRMishra2008a,Ram2010a,Ram2011a,Ram2013a, Ram2013b} as well as by other groups \cite{JunminWang2008a,Yan2006a,Dimova2007a,Swanson1998a}. The use of a red-detuned (from cooling transition) push laser beam of Gaussian transverse profile is a very convenient method to transfer atoms from VC-MOT to UHV-MOT. Such a push beam ejects atoms from VC-MOT and provides transverse confinement (i.e. guiding) to atoms during their transfer from VC-MOT to UHV-MOT region. This guiding improves the number of atoms trapped in the UHV-MOT as more atoms reach the UHV-MOT volume. In our experiments, a laser beam having $\sim$21 mW of power (before entry into the VC-MOT chamber) and detuning $\sim$ 1 GHz, focused to a $1/e^2$ radius of $\sim$ 35 $\mu$m on VC-MOT atom cloud, is used as a push beam for the above atom transfer purpose. During the UHV-MOT loading, UHV-MOT beams are kept on and appropriate current ($\sim$ 0.6-1.0 A) is passed through the UHV-MOT coils to generate the necessary field gradient for the MOT formation. In our experiments, the UHV-MOT is loaded for $\sim$ 40 s duration over which the push beam is kept on. Nearly $2-3\times 10^8$ atoms are obtained in the saturated UHV-MOT in our setup. The temperature of atom cloud in the UHV-MOT ranges from 100 $\mu$K to 400 $\mu$K, depending upon the various parameters set in the experiments. The number and temperature of the atom clouds in both the MOTs are estimated using fluorescence imaging and free-expansion techniques \cite{SRMishra2008a, Ram2013b}

(c) UHV-MOT compression and cooling in molasses:

After fully loading of UHV-MOT, the UHV-MOT atom cloud is kept in a compressed-MOT for $\sim$20 ms duration. This is achieved by increasing the detuning of cooling laser beams for UHV-MOT from $-12 MHz to -25 MHz$ and magnetic field gradient from $6 G/cm $ to $\sim 8 G/cm $. The compressed MOT stage results in a higher density in the MOT cloud. At the end of the compressed MOT stage, the current in the UHV-MOT coils is switched-off and atoms are cooled in optical molasses for $\sim$ 5 ms duration with reduced power (to $20\%$ of original power) and increased detuning ($\sim$ 42 MHz) of UHV-MOT cooling beams. After the optical molasses stage the temperature of atom cloud is obtained in the range of $\sim$40-80 $\mu$K.

(d) Optical pumping:

The UHV-MOT atom cloud after the molasses stage is optically pumped to $F = 2, m_F = 2$ Zeeman hyperfine sublevel of $^{87}$Rb atom for magnetic trapping. This leads to an efficient transfer of atoms from molasses to magnetic trap. The optical pumping is accomplished in $\sim$ 0.5 ms duration. For optical pumping, the atom cloud after molasses is exposed to a weak laser (with $\sigma^+$ polarization) pulse of duration $\sim$0.5 ms in presence of an uniform magnetic field of $\sim$2 G. In order to optimize the optical pumping process, the power in optical pumping pulse can be varied (while keeping the pulse duration fixed at $\sim$ 0.5 ms) and number of atoms in the magnetic trap can be monitored. 

\begin{figure}
	\centering
		\includegraphics[width = 0.8\columnwidth]{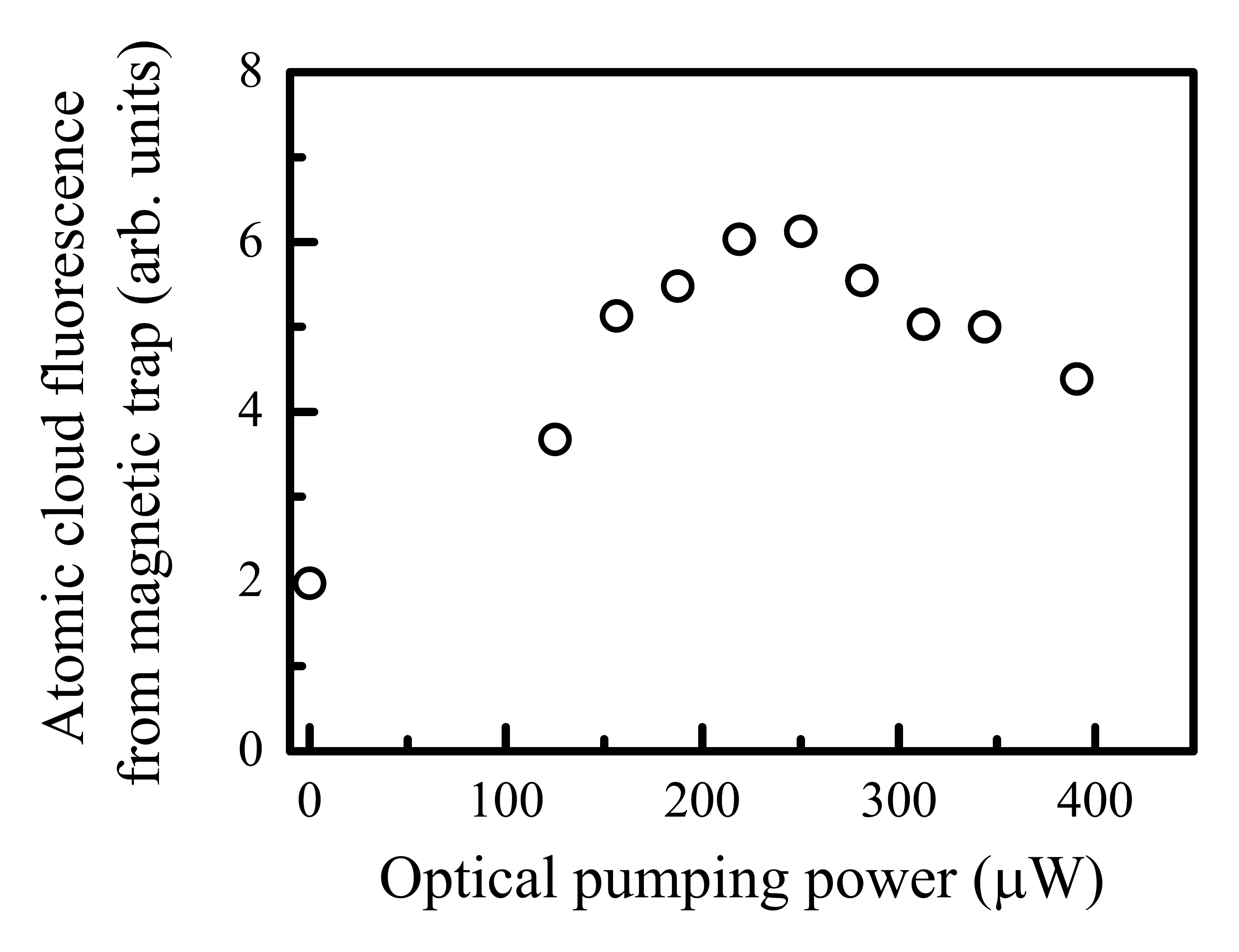}
	\caption{Variation in fluorescence power (proportional to number of atoms in the trap) from the atom cloud in the quadrupole trap as a function of power of the optical pumping beam.}
	\label{fig:optical_pumping_optimization}
\end{figure}

Fig.\ref{fig:optical_pumping_optimization} shows the variation in number of atoms captured in the quadrupole trap for different optical pumping power values. At optimized optical pumping power, we could transfer $\sim$30-40 $\% $ of atoms from UHV-MOT to quadrupole magnetic trap. This can be further improved by improving the optical pumping in our setup and by fast rise of current in the of quadrupole trap coils. At present, the optical pumping field is not a perfect rectangular pulse in shape and quadrupole magnetic field switching is also slow (rise-time $\sim$2.5 ms) 

(e) Magnetic trapping:

\begin{figure}[h]
	\centering
		\includegraphics[width = 0.9\columnwidth]{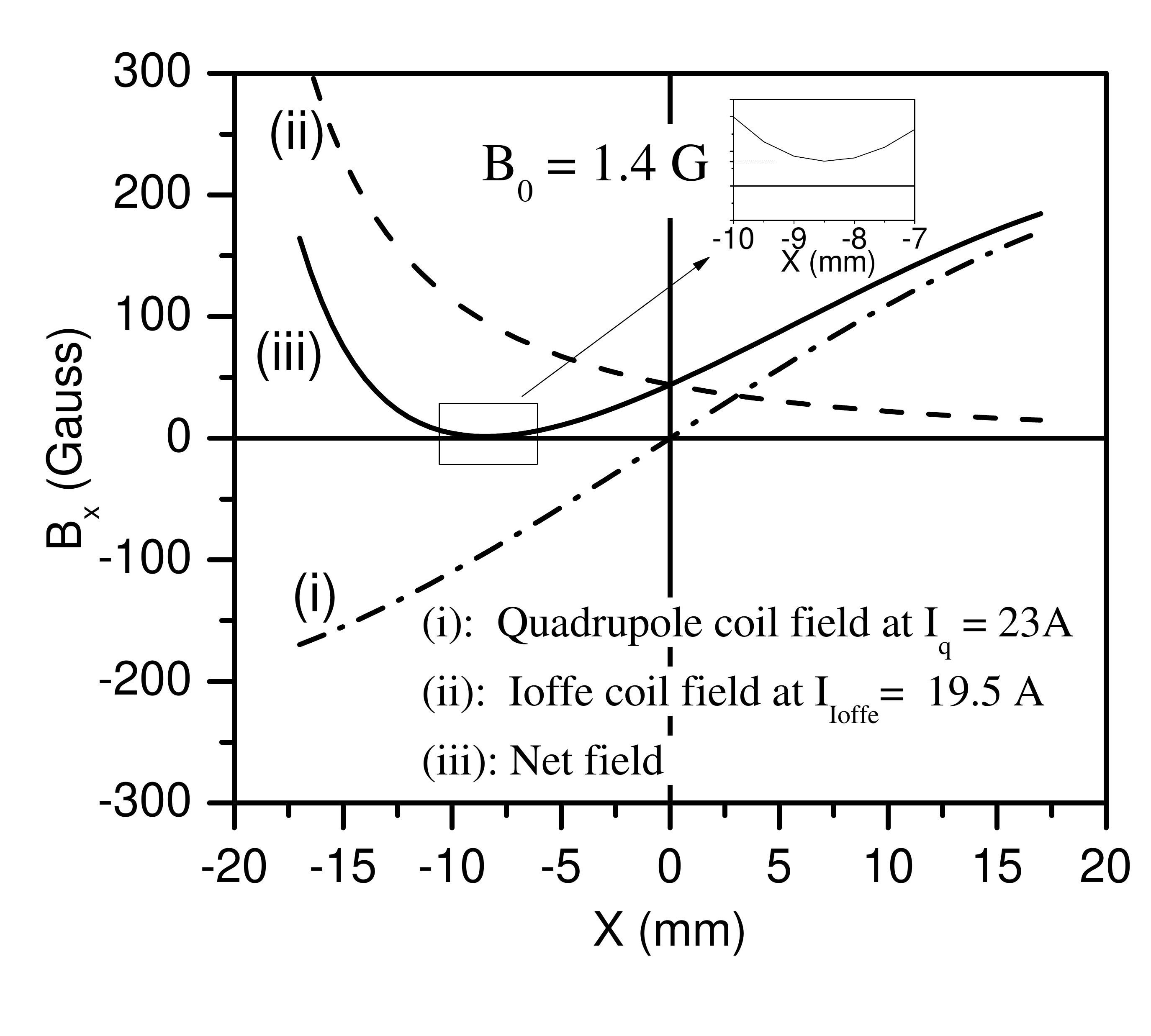}
	\caption{Magnetic field distribution due to different coils along the Ioffe coil axis. The inset shows the non-zero offset of the QUIC trap.}
	\label{fig:Fig5_QUIC_field}
\end{figure}

The UHV-MOT coils are used to form the quadrupole magnetic trap. The quadrupole coils and an Ioffe coil form the desired QUIC trap when currents in these coils is set appropriately. For our QUIC trap, the Ioffe coil has 238 number of turns (with 78 number of turns in the conical part and 160 number of turns in the cylindrical part). The conical section of the Ioffe coil has a length of 16 mm with gradually increasing diameter (from 6 mm to 35 mm) and the cylindrical section is of length 18 mm with diameter 35 mm. The wire used in this coil has diameter $\sim $ 1.0 mm. This specific design of Ioffe coil has been chosen to make approach of the coil to the glass cell without blocking the UHV-MOT beams. The calculated fields due to quadrupole and Ioffe coils are shown in Fig. \ref{fig:Fig5_QUIC_field} for the geometrical configuration as shown in Fig. \ref{fig:QUIC_trap_coil}. From the graphs in Fig. \ref{fig:Fig5_QUIC_field}, it is evident that Ioffe configuration in the QUIC trap is formed with the currents of 23 A and 19.5 A in quadrupole and Ioffe coils respectively. Our measured field values are close to the calculated field values. We used these values during the experiments for QUIC trap formation.

The trapping potential of the QUIC trap near the field minimum position $x_0$ can be approximately given as \cite{Schaff2011a, Meyrath2005a, Yoon2009a},
\begin{equation}
U(r)=\mu |B(\textbf{r})| = \mu B_0 +\frac{m}{2}\left(\omega_{||}^2(x-x_0)^2+\omega_{\perp}^2(y^2+z^2)\right),
\label{eq:QUIC_potential}
\end{equation}
where $m$ is the mass of the atom, $\mu$ is magnetic dipole moment of the atom and $B_0$ is non-zero value of the field at the  minimum of the trapping potential at $x_0$. The frequencies $\omega_{||}=\sqrt{\frac{\mu B_x^{''}}{m}}$ and $\omega_{\perp}=\sqrt{\frac{\mu B_\perp^{'2}}{mB_0}}$ are axial and radial trap frequencies, with $B_x^{''}$ is the curvature of the field along x-axis (i.e. Ioffe coil axis) and $B_{\perp}^{'}$ is the geometric mean of the quadrupole trap field gradients $B_y'$ and $B_z'$ \textit{i.e.}  $B_{\perp}^{'}$=$\sqrt{B_y' B_z'}$. For ($F = 2, m_F = 2$) Zeeman hyperfine sublevel of $^{87}$Rb atom, $\sqrt{\frac{\mu}{m}}=2\pi1.2765$, which gives axial and radial trap frequencies for our QUIC trap as $\omega_{||}(Hz)=2\pi 1.2765 \sqrt{B_x^{''} (G/cm^2)}$ and $\omega_{\perp}(Hz)=2\pi 1.2765\sqrt{\frac{B_\perp^{'2}}{B_0}(G/cm^2)}$ respectively.

\begin{figure}[t]
	\centering
		\includegraphics[width = 0.8\columnwidth]{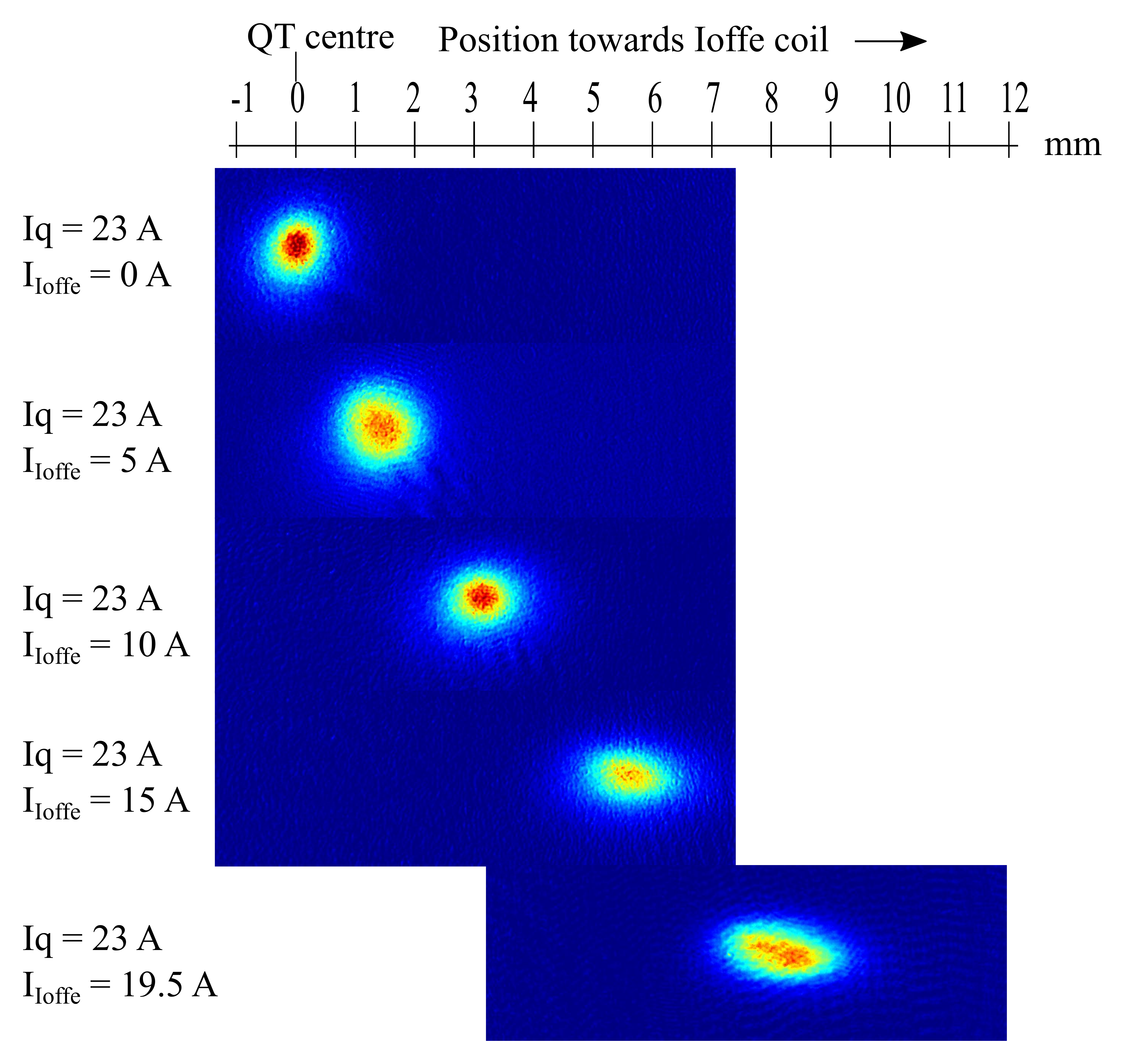}
	\caption{(Color online) Shifting of the atom cloud with the increase in the Ioffe coil current. For the last image in the figure, the imaging system was adjusted to bring the cloud image in the centre of the CCD camera. The colors from blue to red in the OD images show OD values in the increasing order.}
	\label{fig:Fig6_QUIC_formation}
\end{figure}

In the experiments, after optical pumping of atom cloud to ($F = 2, m_F = 2 $) state, the magnetic trapping of atoms is initiated by rapidly (in $\sim$2.5 ms) switching-ON a current of $\sim$13 A in the quadrupole coils. Then current in quadrupole is slowly ramped-up to a final value $I_q$ ($\sim$23 A) in a duration of $\sim$1000 ms for an adiabatic loading of the quadrupole trap. After this, the current in Ioffe coil is ramped-up from 0 to a final value $I_{Ioffe}$ in $\sim$2500 ms to convert quadrupole trap into the QUIC trap. During the conversion of quadrupole trap into QUIC trap, depending upon the current in Ioffe coil, the minimum of the magnetic field (and potential) gets shifted from the quadrupole trap centre (x = 0) to a new position towards the Ioffe coil (-ve side of x-axis) (as shown in Fig. \ref{fig:Fig6_QUIC_formation}). This results in the shifting of trapped atom cloud towards the Ioffe coil, which is dependent on current in the Ioffe coil. Fig. \ref{fig:Fig6_QUIC_formation}  shows this observed shift of atom cloud in our experiments for different values of the current $I_{Ioffe}$ in the Ioffe coil. For these measurements, the value of $I_{Ioffe}$ was changed in the vi of labVIEW and experiment was repeated to capture image for the chosen value of $I_{Ioffe}$. The observed shift of the atom cloud was $\sim$8.5 mm for $I_{Ioffe}$ = 19.5 A (with $I_q$ = 23 A) which is in good agreement with the simulated value of the shift in the minimum of trapping potential as shown in Fig. \ref{fig:Fig5_QUIC_field}. This evidently indicates the appropriate working of QUIC trap in our experiments. The trap frequencies for these values of current ($I_{Ioffe}$ = 19.5 A and $I_q$ = 23 A) are $\omega_{||}=2\pi$. 17.6 $Hz$ and $\omega_{\perp}=2\pi$. 174.2 $Hz$ ( with $\omega_{y}=2\pi$. 122.5  $Hz$ and $\omega_{z}=2\pi$. 246  $Hz$ ). Since the position of atom cloud in the final QUIC trap is shifted far away from the quadrupole trap center, the absorption probe beam and imaging system aligned to quadrupole trap center requires re-positioning to capture the image of the cloud in the final QUIC trap (see last image in Fig. \ref{fig:Fig6_QUIC_formation}).

A longer life-time is always helpful for evaporative cooling. There are several factors on which this life-time is dependent. The vacuum of the system plays a crucial role besides the residual light going into the trap volume. During the magnetic trapping of atoms, it is important that all kinds of light emissions going to magnetic trap volume are blocked, and magnetic trapping is performed in a completely dark environment. For this, we use mechanical shutters at different places to block the leaked radiation from various AOMs used for switching of various laser beams. Also, the trapping region is isolated by putting a physical partition on the table. With appropriate darkness and pressure in the glass cell ($5\times10^{-11}$ Torr), we get a life-time of $\sim$ 20 sec for atom cloud in our QUIC trap.

(f) Absorption probe imaging and image analysis:

\begin{figure}[h]
	\centering
		\includegraphics[width = 0.9\columnwidth]{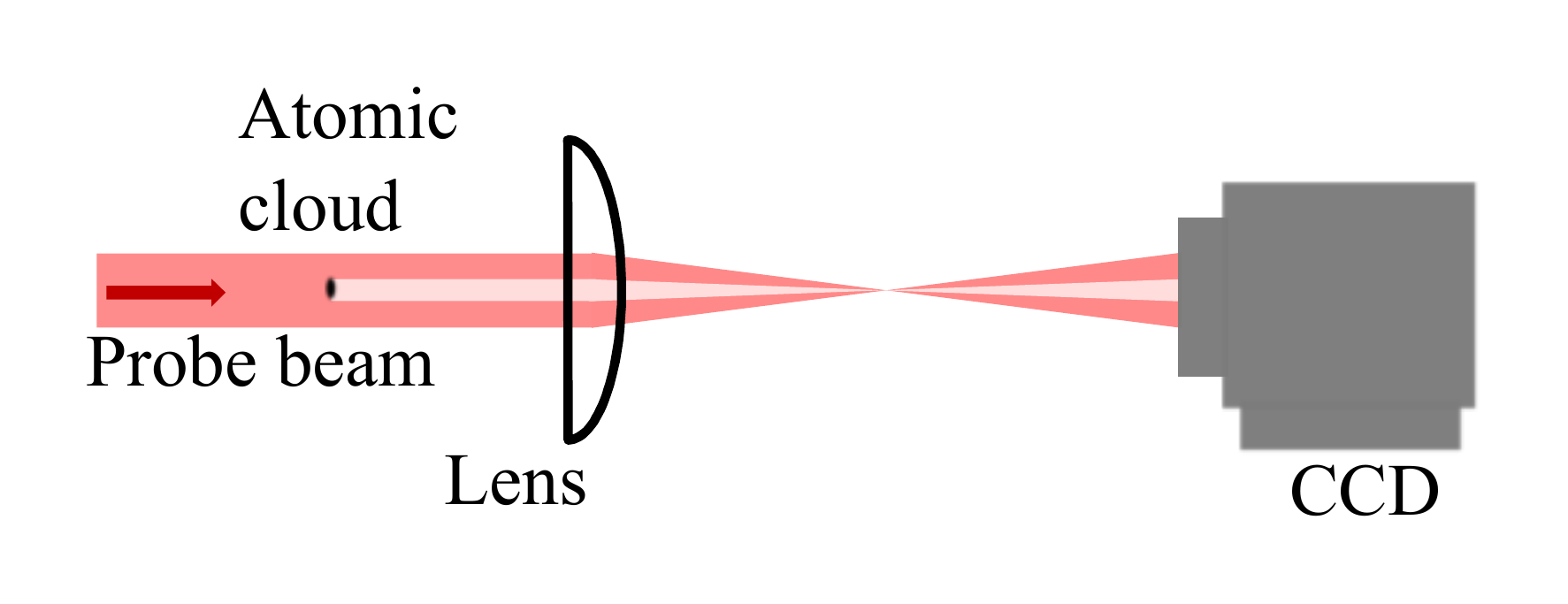}
	\caption{(Color online) The schematic of the absorption probe imaging method.}
	\label{fig:Fig8_AP_schematic}
\end{figure}

The standard absorption probe imaging method has been used to characterize atom cloud in the magnetic trap for number and temperature measurements. In absorption probe technique, the atom cloud is illuminated by a low intensity resonant probe laser beam (propagating along z-axis in our case) and the absorptive shadow of the atom cloud in the beam is imaged onto a charged coupled device (CCD) camera using an appropriate optical imaging system as shown schematically in Fig. \ref{fig:Fig8_AP_schematic}. In the experiments, to implement the absorption probe imaging technique, we grab three images to get the required information; (i) the background image which is obtained without absorption probe beam ($IBG$), (ii) the image of the probe beam without atom cloud ($IP$), and (iii) the image of the probe beam transmitted through the atom cloud ($IT$). From these three images and using an image processing program, we construct the optical density ($OD$) image of the atom cloud using the following equation,

\begin{equation}
OD(i,j) = \ln \frac{IP(i, j) - IBG(i, j)}{IT(i, j) - IBG(i, j)}
\end{equation}
where $IBG(i,j)$, $IP(i, j)$ and $IT(i, j)$ denote the CCD counts at pixel $(i,j)$ in $IBG$ , $IP$ and $IT$ images respectively. This $OD$ image gives the column density of the atom cloud as $OD(x,y) = \int (\sigma_0. n(x,y,z) dz)$, where $\sigma_0$ is absorption cross section. The plot of $OD(x,y)$ gives important information related to the transverse distribution of column density in the cloud.

(g) RF evaporative cooling:

In radio frequency (RF) evaporative cooling, RF field is used to force the ejection of hotter atoms from the trapped atom cloud so that remaining cloud is settled to lower temperature after thermalization \cite{Petrich1995a,Pritchard1983a}. For evaporative cooling of $^{87}$Rb atoms trapped in the QUIC trap, we apply radio frequency (RF) radiation emitted from a single loop antenna kept at one side of the glass cell with its axis aligned along the quadrupole trap axis (z-axis). We use a programmable synthesizer (Agilent 55332A) as an RF source whose output is connected to an RF amplifier. The antenna is connected to the RF amplifier through an impedance-matching circuit. A logarithmic variation of RF frequency with time (from high initial value to low final value), called RF scan, is implemented to achieve efficient evaporative cooling \cite{Lu2004a}. The useful frequency range for RF scan for RF-evaporation can be roughly estimated by knowing the variation of magnetic field with position in the trap (to know the position dependent Zeeman splitting) and size of the atom cloud in the trap. The process of RF-evaporation is very sensitive to the final frequency in the scan range.

\section{Results and discussion}
\label{results_discussion}

After completion of the RF evaporative cooling by frequency scan of the RF field (i.e. RF scan), the trapped atom cloud OD images were obtained by in-situ absorption imaging using a near resonant probe beam (pulse duration 60$\mu s$) and imaging optics as discussed before. These images of evaporatively cooled cloud are recorded for different ranges of RF scan to find the appropriate scan range for effective evaporative cooling for BEC phase transition. Since imaging measurement heats up the sample, in order to record the image for the changed parameters, the trap is reloaded and evaporative cooling cycle is repeated again. Because of the several steps involved in a cycle of the experiment, the recorded images show some cycle-to-cycle fluctuation in the number of atoms ($\pm$ 10 $\%$) even without changing the experimental parameters. The evaporative cooling experiments are performed over a number of cycles to record the absorption images for different ranges of RF scan. 

\begin{figure}[t]
	\centering
		\includegraphics[width = 0.8\columnwidth]{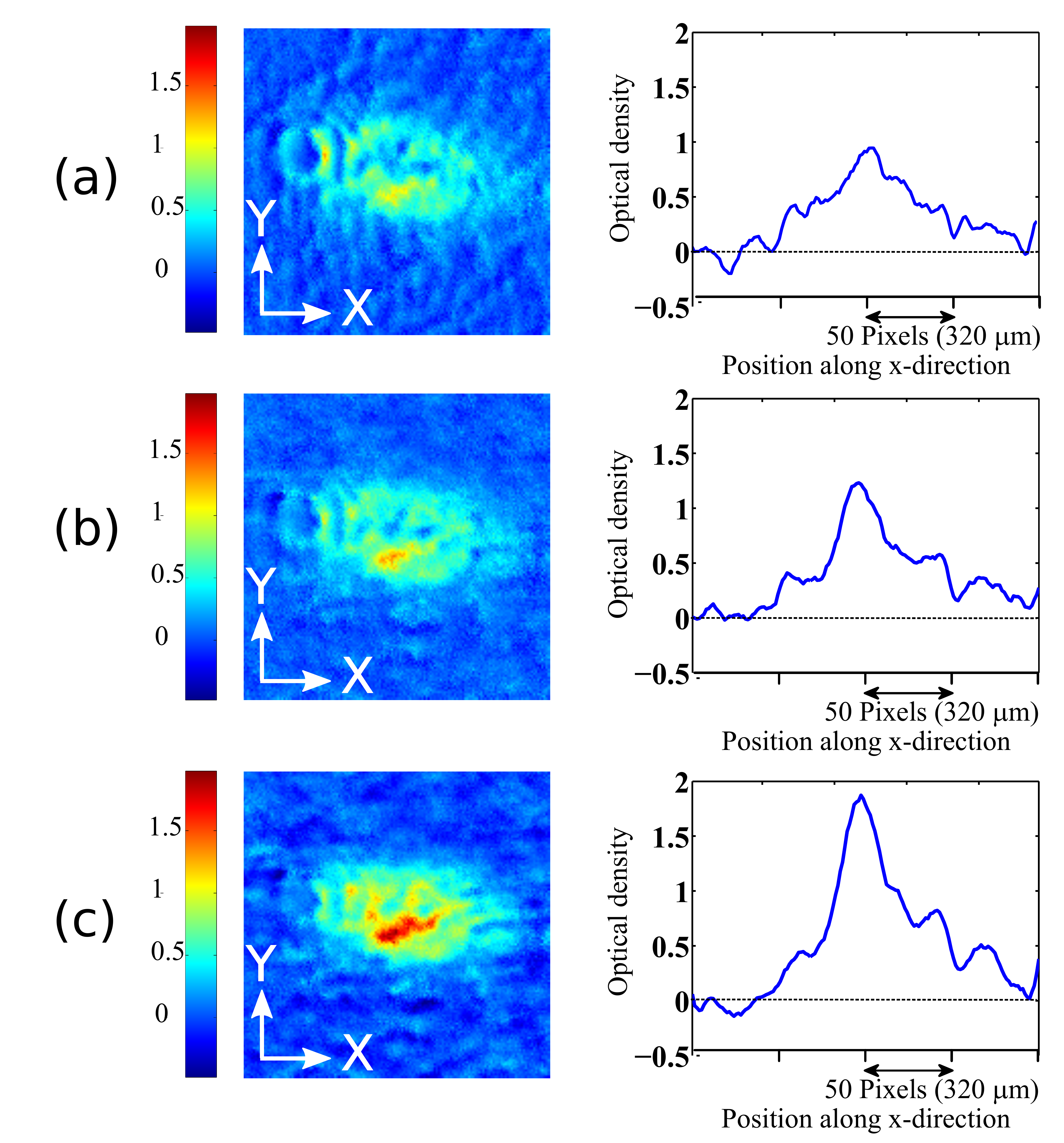}
	\caption{(Color online) Images and spatial profiles of the atom cloud cooled by RF evaporation for different scan ranges of frequency. (a) : 20 MHz to 2.5 MHz, (b) : 20 MHz to 2.3 MHz and (c) : 20 MHz to 2.2 MHz. The colors from blue to red in the OD images show OD values in the increasing order.}
	\label{fig:Fig9_bimodal}
\end{figure}

Fig. \ref{fig:Fig9_bimodal} shows the OD images and spatial profiles of the atom cloud cooled by using different ranges of RF scan. The power of RF-field was $\sim$6-8 W over these ranges. For some of RF scan ranges, the spatial profile of OD becomes bimodal (as shown in Fig. \ref{fig:Fig9_bimodal}), with a sharp peak at the centre against a broad Gaussian distribution at the skirts. This bimodal distribution indicates the on-set of Bose-Einstein condensation of atoms at the trap centre. This kind of bimodal distribution in OD profile is a well known signature of occurrence of BEC during the evaporative cooling experiments \cite{Ketterle1999a}. We note that the bimodal spatial profile of OD is observable only for appropriately chosen parameters during the RF evaporation experiments. The appearance of condensate part in the bimodal OD profile after RF evaporation is quite sensitive to the final value of frequency in the RF scan.

\begin{figure}
	\centering
		\includegraphics[width=0.5\columnwidth]{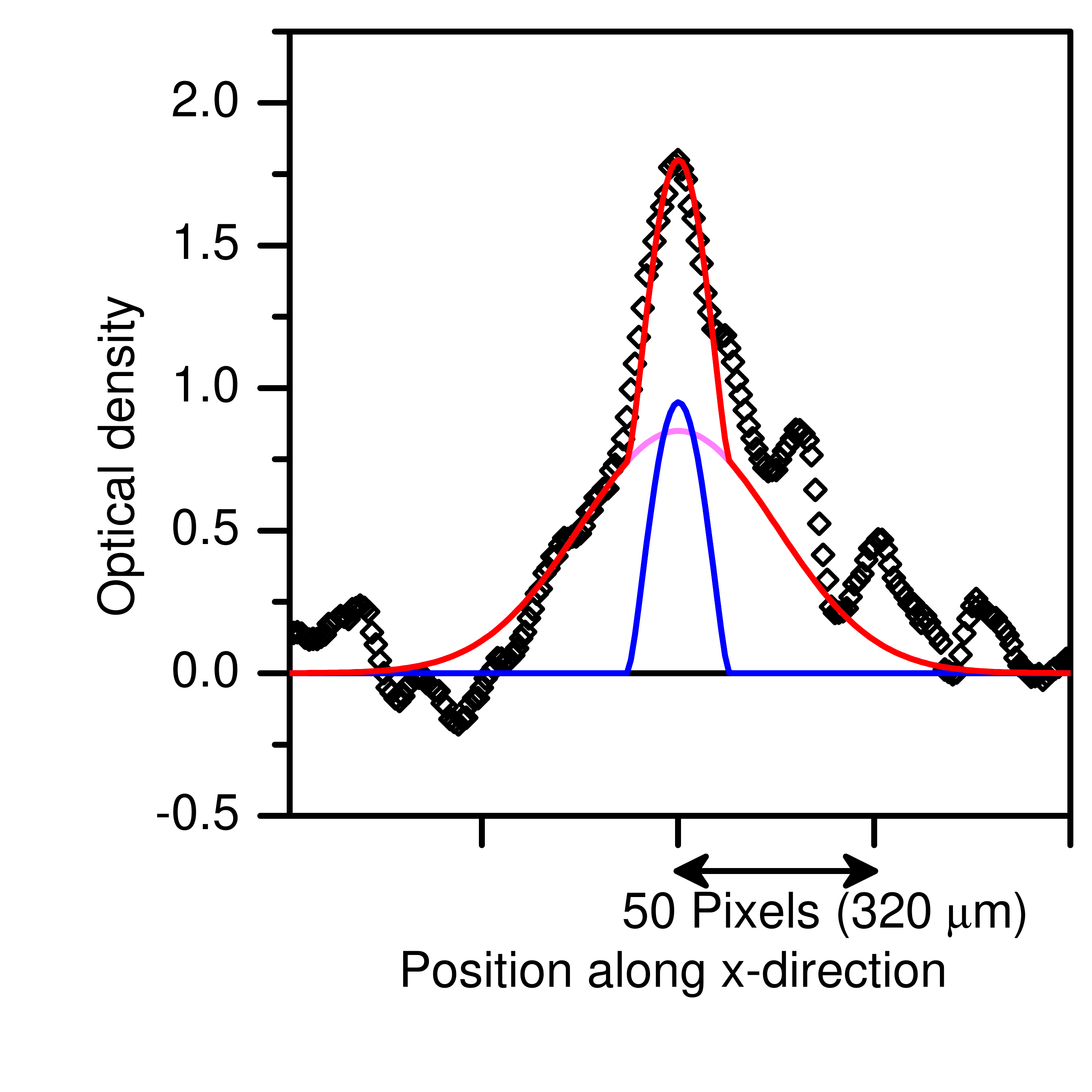}
	\caption{ (Color online) Measured optical density ($OD$) profile (diamonds) along x-axis and its fits to Eq. (\ref{eq:bimodal}) for thermal (pink curve) and Bose condensate (blue curve) parts. The red curve shows the resultant fit.}
	\label{fig:Fitted_profile}
\end{figure}

 The total number $N$ can be estimated by summing the column density values in the recorded OD image. The bimodal profile for the column density can be written as \cite{Stamper-Kurn1998a}, 
 
 \begin{equation}
 \begin{split}
 	\tilde{n}(x,y) = \tilde{n}_{c} \max\left({1-\left(\frac{x-x_0}{w_x/2}\right)^2}-\left(\frac{y-y_0}{w_y/2}\right)^2,0\right)^{3/2}+ \\
 	\tilde{n}_{th} \exp \left[-\frac{(x-x_0)^2}{2(\sigma_x/2)^2}-\frac{(y-y_0)^2}{2(\sigma_y/2)^2}\right],
\end{split}
\label{eq:bimodal}
\end{equation}
where $\tilde{n}_{th}$ and $\tilde{n}_c$ are peak column density values for thermal and condensate clouds and $\sigma_x, \sigma_y$, $w_x$ and $w_y$ are width parameters for the thermal and condensate parts of the cloud. The temperature ($T$) and chemical potential ($\mu$) can be estimated by fitting the observed OD profile to Eq. (\ref{eq:bimodal}) and using the relations $k_BT =\frac{1}{4} m\omega_x^2\sigma_x^2$  and $\mu = \frac{1}{8}m\omega_x^2w_x^2$ as discussed in ref. \cite{Stamper-Kurn1998a}, where $k_B$ is Boltzmann's constant. A typical fit is shown in Fig. \ref{fig:Fitted_profile}. For an image shown in Fig. \ref{fig:Fig9_bimodal}(c), these parameters are  $T \sim 3.2$ $\mu K$, $\mu \sim 440$ $nK$ and $N \sim 6\times 10^5$.

\begin{figure}[h]
	\centering
	\includegraphics[width = \columnwidth]{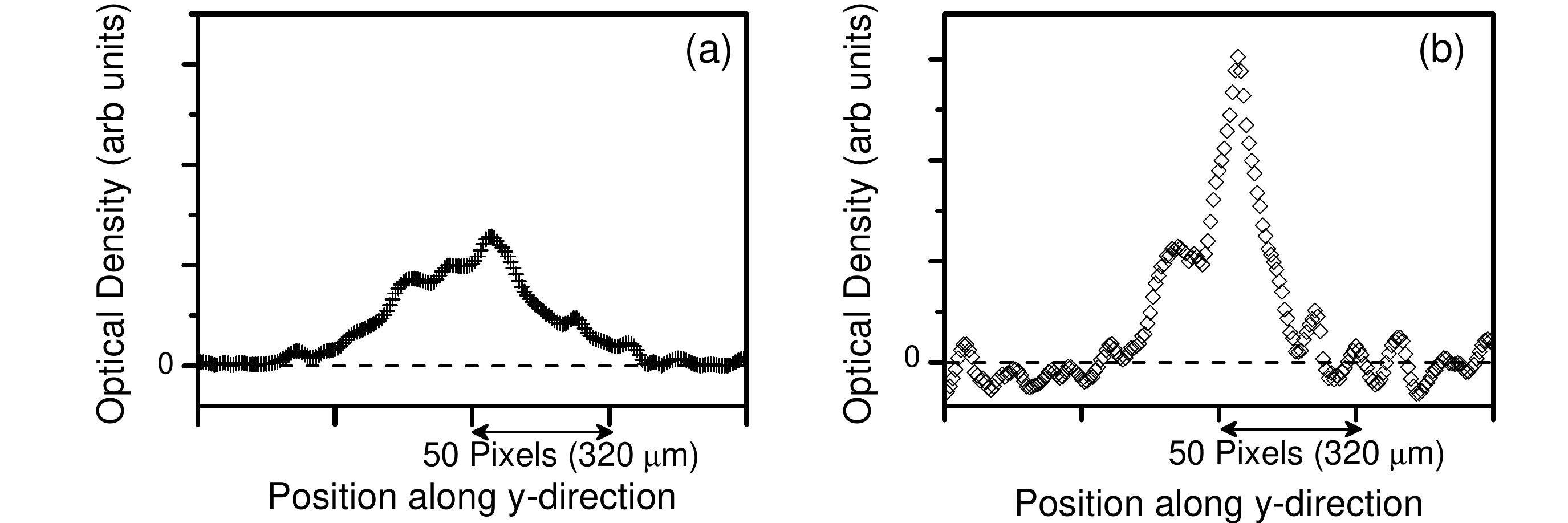}
	\caption{Optical density profile along y-direction. (a) pure thermal cloud  and (b) thermal cloud with condensate. The profile is shown in y-direction in which condensate is tightly confined in the QUIC trap. The profile in (b) shows more negative $OD$ values (i.e. presence of halo) as compared to the profile in (a).}
		\label{fig:Fig11_negative_OD}
\end{figure}

In the OD images in which bimodal distribution is prominent, a halo surrounding the peak of OD is observed. This halo region has negative OD value (Fig. \ref{fig:Fig11_negative_OD}), which implies a higher light intensity in this region in the probe beam image with atom cloud than the intensity in the probe image without atom cloud. Similar halo region has been reported earlier \cite{Bradley1995a,Yu-Zhu2003a} and attributed to the diffraction of probe beam from the dense and spatially localised Bose condensate cloud \cite{Bradley1995a}. The $OD$ image profiles which do not show bimodal distribution, a negligible negative $OD$ is observed (as evident from Fig. \ref{fig:Fig11_negative_OD}). 

\begin{figure}
	\centering
		\includegraphics[width = \columnwidth ]{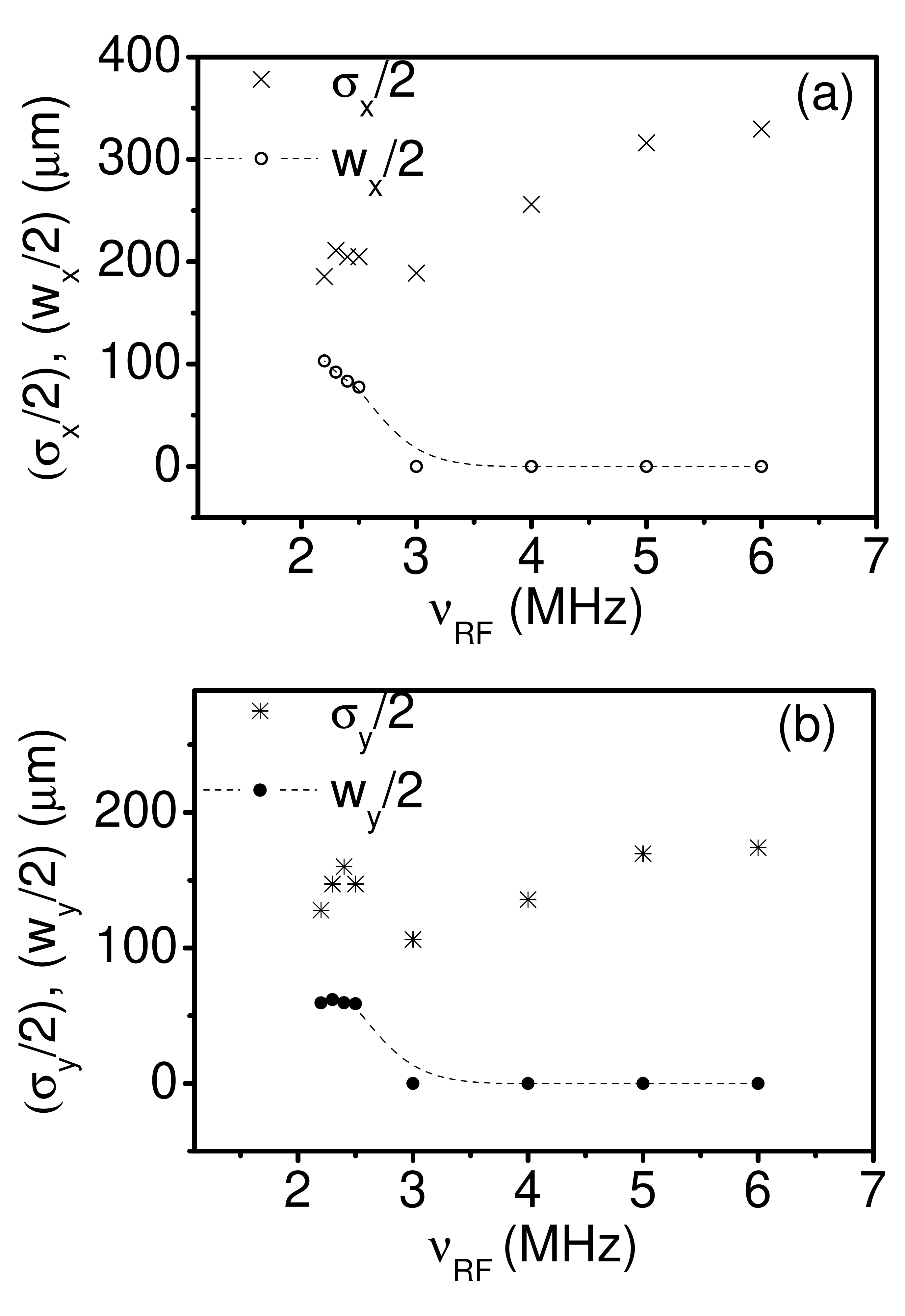}
	\caption{Measured variation in the cloud sizes along x- and y- directions with final frequency ($\nu_{RF}$) in the RF scan. The scan range was from 20 MHz to $\nu_{RF}$. }
	\label{fig:sigma_xy}
\end{figure}

It is known that the size of cloud in the magnetic trap is dependent on its temperature \cite{Stamper-Kurn1998a,Zhang2005a}. Since the temperature of evaporatively cooled atom cloud is also dependent on the final value of RF frequency in the RF scan \cite{Stamper-Kurn1998a, Yu-Zhu2003a}, we attempted to study the variation in the cooled atom cloud size with the value of final frequency in the RF scan. Fig. \ref{fig:sigma_xy} shows the observed variation in the sizes of the thermal and condensate clouds with the value of final value of frequency in the RF scan. Here sizes were estimated by fitting eq. \ref{eq:bimodal} to the measured $OD$ profiles of the cooled atom cloud. An example of the fitting is shown in Fig. \ref{fig:Fitted_profile}. The data in the Fig. \ref{fig:sigma_xy} shows the decrease in the thermal cloud size with decrease in the final frequency, before the appearance of the bimodal $OD$ profile. When final frequency is reduced below a certain value, the $OD$ profile becomes bimodal. In this bimodal distribution, the size of condensate part is much smaller than the thermal part (Fig.\ref{fig:Fig9_bimodal}). This is consistent with the prediction of sudden decrease in the size of the cloud below the critical temperature \cite{Durfee1998a, Zhang2005a}. Fig \ref{fig:sigma_xy}(a) shows that when condensate part appears in the $OD$ profile, the condensate size starts increasing with decrease in the final frequency in the RF scan. This increase in the size can be attributed to the increased repulsive interactions in the condensate with increase in the number of atoms in the condensate, as has been discussed in Ref. \cite{Zhang2005a}. We note that increase in the condensate width with decrease in final frequency is more clearly observable along the weakly confining trap axis (x-axis), as compared to the strongly confining y-axis. Further, in the presence of the condensate, the size of thermal cloud also increases with decrease in the final frequency as we observed in our experiments (Fig. \ref{fig:sigma_xy}). This is consistent with the predictions made earlier \cite{Naraschewski1998a,Zhang2005a} that presence of condensate pushes the thermal cloud out. Thus our size dependent observations shown in Fig. \ref{fig:sigma_xy} are consistent with the predictions made earlier.

\section{Conclusion}
\label{conclusion}
We have presented the generation and characterization of Bose condensate of $^{87}$Rb atoms in our home-made experimental setup. The finally cooled atom cloud in the QUIC trap has been characterized by in-situ absorption imaging technique. The observed bimodal spatial profiles of optical density are suggestive of the presence of the Bose condensate in the trap. We have observed variation in the sizes of thermal and condensate clouds with the final frequency of the RF-field used for evaporative cooling. These observed results on frequency dependent sizes of thermal and condensate clouds are consistent with the theory and predictions made earlier.

\section*{Acknowledgement}
We thank V. B. Tiwari, S. Singh, V. Singh, A. Srivastava and A. Chakraborty for their help in the experiments. We also thankful to V. Bhanage, P. P. Deshpande, S. Tiwari, L. Jain, and A. Pathak for developing controller system, C. Rajan and P. Kumar for the development of switching circuitry, M. Lad and  P. S. Bagduwal for providing RF amplifier, H. S. Vora for image processing software, K. V. A. N. P. S. Kumar and S. K. Shukla for help in UHV system.


\begin{thebibliography}{48}
\expandafter\ifx\csname natexlab\endcsname\relax\def\natexlab#1{#1}\fi
\expandafter\ifx\csname bibnamefont\endcsname\relax
  \def\bibnamefont#1{#1}\fi
\expandafter\ifx\csname bibfnamefont\endcsname\relax
  \def\bibfnamefont#1{#1}\fi
\expandafter\ifx\csname citenamefont\endcsname\relax
  \def\citenamefont#1{#1}\fi
\expandafter\ifx\csname url\endcsname\relax
  \def\url#1{\texttt{#1}}\fi
\expandafter\ifx\csname urlprefix\endcsname\relax\def\urlprefix{URL }\fi
\providecommand{\bibinfo}[2]{#2}
\providecommand{\eprint}[2][]{\url{#2}}

\bibitem[{\citenamefont{Adams and Riis}(1997)}]{Adams1997a}
\bibinfo{author}{\bibfnamefont{C.}~\bibnamefont{Adams}} \bibnamefont{and}
  \bibinfo{author}{\bibfnamefont{E.}~\bibnamefont{Riis}},
  \bibinfo{journal}{Prog. Quant. Electr.} \textbf{\bibinfo{volume}{21}},
  \bibinfo{pages}{1 } (\bibinfo{year}{1997}), ISSN \bibinfo{issn}{0079-6727},
  \urlprefix\url{http://www.sciencedirect.com/science/article/pii/S0079672796000067}.

\bibitem[{\citenamefont{Anderson
  et~al.}(1995{\natexlab{a}})\citenamefont{Anderson, Ensher, Matthews, Wieman,
  and Cornell}}]{Anderson1995a}
\bibinfo{author}{\bibfnamefont{M.~H.} \bibnamefont{Anderson}},
  \bibinfo{author}{\bibfnamefont{J.~R.} \bibnamefont{Ensher}},
  \bibinfo{author}{\bibfnamefont{M.~R.} \bibnamefont{Matthews}},
  \bibinfo{author}{\bibfnamefont{C.~E.} \bibnamefont{Wieman}},
  \bibnamefont{and} \bibinfo{author}{\bibfnamefont{E.~A.}
  \bibnamefont{Cornell}}, \bibinfo{journal}{Science}
  \textbf{\bibinfo{volume}{269}}, \bibinfo{pages}{198}
  (\bibinfo{year}{1995}{\natexlab{a}}),
  \eprint{http://www.sciencemag.org/content/269/5221/198.full.pdf},
  \urlprefix\url{http://www.sciencemag.org/content/269/5221/198.abstract}.

\bibitem[{Ing(1999)}]{Inguscio1999a}
in \emph{\bibinfo{booktitle}{Proceedings of the International School of
  Physics, {E}nrico {F}ermi Course {CXL}}}, edited by
  \bibinfo{editor}{\bibfnamefont{M.}~\bibnamefont{Inguscio}},
  \bibinfo{editor}{\bibfnamefont{S.}~\bibnamefont{Stringari}},
  \bibnamefont{and} \bibinfo{editor}{\bibfnamefont{C.~E.} \bibnamefont{Wieman}}
  (\bibinfo{publisher}{IOS Press, Amsterdam}, \bibinfo{year}{1999}).

\bibitem[{\citenamefont{Dwyer et~al.}(2005)\citenamefont{Dwyer, Gay,
  de~Lesegno, Weiner, Camposeo, Tantussi, Fuso, Allegrini, and
  Arimondo}}]{Dwyer2005a}
\bibinfo{author}{\bibfnamefont{C.~O.} \bibnamefont{Dwyer}},
  \bibinfo{author}{\bibfnamefont{G.}~\bibnamefont{Gay}},
  \bibinfo{author}{\bibfnamefont{B.~V.} \bibnamefont{de~Lesegno}},
  \bibinfo{author}{\bibfnamefont{J.}~\bibnamefont{Weiner}},
  \bibinfo{author}{\bibfnamefont{A.}~\bibnamefont{Camposeo}},
  \bibinfo{author}{\bibfnamefont{F.}~\bibnamefont{Tantussi}},
  \bibinfo{author}{\bibfnamefont{F.}~\bibnamefont{Fuso}},
  \bibinfo{author}{\bibfnamefont{M.}~\bibnamefont{Allegrini}},
  \bibnamefont{and} \bibinfo{author}{\bibfnamefont{E.}~\bibnamefont{Arimondo}},
  \bibinfo{journal}{Nanotechnology} \textbf{\bibinfo{volume}{16}},
  \bibinfo{pages}{1536} (\bibinfo{year}{2005}),
  \urlprefix\url{http://stacks.iop.org/0957-4484/16/i=9/a=022}.

\bibitem[{\citenamefont{Ludlow et~al.}(2015)\citenamefont{Ludlow, Boyd, Ye,
  Peik, and Schmidt}}]{Ludlow2015}
\bibinfo{author}{\bibfnamefont{A.~D.} \bibnamefont{Ludlow}},
  \bibinfo{author}{\bibfnamefont{M.~M.} \bibnamefont{Boyd}},
  \bibinfo{author}{\bibfnamefont{J.}~\bibnamefont{Ye}},
  \bibinfo{author}{\bibfnamefont{E.}~\bibnamefont{Peik}}, \bibnamefont{and}
  \bibinfo{author}{\bibfnamefont{P.~O.} \bibnamefont{Schmidt}},
  \bibinfo{journal}{Rev. Mod. Phys.} \textbf{\bibinfo{volume}{87}},
  \bibinfo{pages}{637} (\bibinfo{year}{2015}),
  \urlprefix\url{http://link.aps.org/doi/10.1103/RevModPhys.87.637}.

\bibitem[{\citenamefont{Gauguet et~al.}(2009)\citenamefont{Gauguet, Canuel,
  L\'ev\`eque, Chaibi, and Landragin}}]{Gauguet2009a}
\bibinfo{author}{\bibfnamefont{A.}~\bibnamefont{Gauguet}},
  \bibinfo{author}{\bibfnamefont{B.}~\bibnamefont{Canuel}},
  \bibinfo{author}{\bibfnamefont{T.}~\bibnamefont{L\'ev\`eque}},
  \bibinfo{author}{\bibfnamefont{W.}~\bibnamefont{Chaibi}}, \bibnamefont{and}
  \bibinfo{author}{\bibfnamefont{A.}~\bibnamefont{Landragin}},
  \bibinfo{journal}{Phys. Rev. A} \textbf{\bibinfo{volume}{80}},
  \bibinfo{pages}{063604} (\bibinfo{year}{2009}),
  \urlprefix\url{http://link.aps.org/doi/10.1103/PhysRevA.80.063604}.

\bibitem[{\citenamefont{Fang and Qin}(2012)}]{FangQin2012}
\bibinfo{author}{\bibfnamefont{J.}~\bibnamefont{Fang}} \bibnamefont{and}
  \bibinfo{author}{\bibfnamefont{J.}~\bibnamefont{Qin}},
  \bibinfo{journal}{Sensors} \textbf{\bibinfo{volume}{12}},
  \bibinfo{pages}{6331} (\bibinfo{year}{2012}), ISSN \bibinfo{issn}{1424-8220},
  \urlprefix\url{http://www.mdpi.com/1424-8220/12/5/6331}.

\bibitem[{\citenamefont{Bodart et~al.}(2010)\citenamefont{Bodart, Merlet,
  Malossi, Dos~Santos, Bouyer, and Landragin}}]{Bodart2010gravimeter}
\bibinfo{author}{\bibfnamefont{Q.}~\bibnamefont{Bodart}},
  \bibinfo{author}{\bibfnamefont{S.}~\bibnamefont{Merlet}},
  \bibinfo{author}{\bibfnamefont{N.}~\bibnamefont{Malossi}},
  \bibinfo{author}{\bibfnamefont{F.~P.} \bibnamefont{Dos~Santos}},
  \bibinfo{author}{\bibfnamefont{P.}~\bibnamefont{Bouyer}}, \bibnamefont{and}
  \bibinfo{author}{\bibfnamefont{A.}~\bibnamefont{Landragin}},
  \bibinfo{journal}{Applied Physics Letters} \textbf{\bibinfo{volume}{96}},
  \bibinfo{eid}{134101} (\bibinfo{year}{2010}),
  \urlprefix\url{http://scitation.aip.org/content/aip/journal/apl/96/13/10.1063/1.3373917}.

\bibitem[{\citenamefont{Behbood et~al.}(2013)\citenamefont{Behbood,
  Martin~Ciurana, Colangelo, Napolitano, Mitchell, and Sewell}}]{Behbood2013}
\bibinfo{author}{\bibfnamefont{N.}~\bibnamefont{Behbood}},
  \bibinfo{author}{\bibfnamefont{F.}~\bibnamefont{Martin~Ciurana}},
  \bibinfo{author}{\bibfnamefont{G.}~\bibnamefont{Colangelo}},
  \bibinfo{author}{\bibfnamefont{M.}~\bibnamefont{Napolitano}},
  \bibinfo{author}{\bibfnamefont{M.~W.} \bibnamefont{Mitchell}},
  \bibnamefont{and} \bibinfo{author}{\bibfnamefont{R.~J.}
  \bibnamefont{Sewell}}, \bibinfo{journal}{Applied Physics Letters}
  \textbf{\bibinfo{volume}{102}}, \bibinfo{eid}{173504} (\bibinfo{year}{2013}),
  \urlprefix\url{http://scitation.aip.org/content/aip/journal/apl/102/17/10.1063/1.4803684}.

\bibitem[{\citenamefont{Lewenstein et~al.}(2007)\citenamefont{Lewenstein,
  Sanpera, Ahufinger, Damski, Sen(De), and Sen}}]{Lewenstein2007a}
\bibinfo{author}{\bibfnamefont{M.}~\bibnamefont{Lewenstein}},
  \bibinfo{author}{\bibfnamefont{A.}~\bibnamefont{Sanpera}},
  \bibinfo{author}{\bibfnamefont{V.}~\bibnamefont{Ahufinger}},
  \bibinfo{author}{\bibfnamefont{B.}~\bibnamefont{Damski}},
  \bibinfo{author}{\bibfnamefont{A.}~\bibnamefont{Sen(De)}}, \bibnamefont{and}
  \bibinfo{author}{\bibfnamefont{U.}~\bibnamefont{Sen}},
  \bibinfo{journal}{Advances in Physics} \textbf{\bibinfo{volume}{56}},
  \bibinfo{pages}{243} (\bibinfo{year}{2007}),
  \eprint{http://dx.doi.org/10.1080/00018730701223200},
  \urlprefix\url{http://dx.doi.org/10.1080/00018730701223200}.

\bibitem[{\citenamefont{Inguscio and Fallani}(2013)}]{Inguscio2013}
\bibinfo{author}{\bibfnamefont{M.}~\bibnamefont{Inguscio}} \bibnamefont{and}
  \bibinfo{author}{\bibfnamefont{L.}~\bibnamefont{Fallani}},
  \emph{\bibinfo{title}{Atomic Physics: Precise Measurements and Ultracold
  Matter}} (\bibinfo{publisher}{Oxford University Press Inc. New York},
  \bibinfo{year}{2013}).

\bibitem[{\citenamefont{Morsch and Oberthaler}(2006)}]{Morsch2006}
\bibinfo{author}{\bibfnamefont{O.}~\bibnamefont{Morsch}} \bibnamefont{and}
  \bibinfo{author}{\bibfnamefont{M.}~\bibnamefont{Oberthaler}},
  \bibinfo{journal}{Rev. Mod. Phys.} \textbf{\bibinfo{volume}{78}},
  \bibinfo{pages}{179} (\bibinfo{year}{2006}),
  \urlprefix\url{http://link.aps.org/doi/10.1103/RevModPhys.78.179}.

\bibitem[{\citenamefont{Bloch et~al.}(2008)\citenamefont{Bloch, Dalibard, and
  Zwerger}}]{Bloch2008a}
\bibinfo{author}{\bibfnamefont{I.}~\bibnamefont{Bloch}},
  \bibinfo{author}{\bibfnamefont{J.}~\bibnamefont{Dalibard}}, \bibnamefont{and}
  \bibinfo{author}{\bibfnamefont{W.}~\bibnamefont{Zwerger}},
  \bibinfo{journal}{Reviews of Modern Physics} \textbf{\bibinfo{volume}{80}},
  \bibinfo{pages}{885} (\bibinfo{year}{2008}), \bibinfo{note}{cited By (since
  1996)1892}.

\bibitem[{\citenamefont{Bloch et~al.}(2012)\citenamefont{Bloch, Dalibard, and
  Nascimbene}}]{Bloch2012a}
\bibinfo{author}{\bibfnamefont{I.}~\bibnamefont{Bloch}},
  \bibinfo{author}{\bibfnamefont{J.}~\bibnamefont{Dalibard}}, \bibnamefont{and}
  \bibinfo{author}{\bibfnamefont{S.}~\bibnamefont{Nascimbene}},
  \bibinfo{journal}{Nat Phys} \textbf{\bibinfo{volume}{8}},
  \bibinfo{pages}{267} (\bibinfo{year}{2012}), ISSN \bibinfo{issn}{1745-2473},
  \urlprefix\url{http://dx.doi.org/10.1038/nphys2259}.

\bibitem[{\citenamefont{Inouye et~al.}(1998)\citenamefont{Inouye, Andrews,
  Stenger, Miesner, Stamper-Kurn, and Ketterle}}]{Inouye1998a}
\bibinfo{author}{\bibfnamefont{S.}~\bibnamefont{Inouye}},
  \bibinfo{author}{\bibfnamefont{M.~R.} \bibnamefont{Andrews}},
  \bibinfo{author}{\bibfnamefont{J.}~\bibnamefont{Stenger}},
  \bibinfo{author}{\bibfnamefont{H.-J.} \bibnamefont{Miesner}},
  \bibinfo{author}{\bibfnamefont{D.~M.} \bibnamefont{Stamper-Kurn}},
  \bibnamefont{and} \bibinfo{author}{\bibfnamefont{W.}~\bibnamefont{Ketterle}},
  \bibinfo{journal}{Nature} \textbf{\bibinfo{volume}{392}},
  \bibinfo{pages}{151} (\bibinfo{year}{1998}), ISSN \bibinfo{issn}{0028-0836},
  \urlprefix\url{http://dx.doi.org/10.1038/32354}.

\bibitem[{\citenamefont{Chin et~al.}(2010)\citenamefont{Chin, Grimm, Julienne,
  and Tiesinga}}]{Chin2010a}
\bibinfo{author}{\bibfnamefont{C.}~\bibnamefont{Chin}},
  \bibinfo{author}{\bibfnamefont{R.}~\bibnamefont{Grimm}},
  \bibinfo{author}{\bibfnamefont{P.}~\bibnamefont{Julienne}}, \bibnamefont{and}
  \bibinfo{author}{\bibfnamefont{E.}~\bibnamefont{Tiesinga}},
  \bibinfo{journal}{Rev. Mod. Phys.} \textbf{\bibinfo{volume}{82}},
  \bibinfo{pages}{1225} (\bibinfo{year}{2010}),
  \urlprefix\url{http://link.aps.org/doi/10.1103/RevModPhys.82.1225}.

\bibitem[{\citenamefont{Cohen-Tannoudji and
  Gu\'{e}ry-Odelin}(2011)}]{Cohen-Tannoudji2011a}
\bibinfo{author}{\bibfnamefont{C.}~\bibnamefont{Cohen-Tannoudji}}
  \bibnamefont{and}
  \bibinfo{author}{\bibfnamefont{D.}~\bibnamefont{Gu\'{e}ry-Odelin}},
  \emph{\bibinfo{title}{Advances in atomic physics: an overview}}
  (\bibinfo{publisher}{World Scientific Publishing Co. Pte. Ltd., Singapore},
  \bibinfo{year}{2011}).

\bibitem[{\citenamefont{Zobay and Garraway}(2001)}]{Zobay2001}
\bibinfo{author}{\bibfnamefont{O.}~\bibnamefont{Zobay}} \bibnamefont{and}
  \bibinfo{author}{\bibfnamefont{B.~M.} \bibnamefont{Garraway}},
  \bibinfo{journal}{Phys. Rev. Lett.} \textbf{\bibinfo{volume}{86}},
  \bibinfo{pages}{1195} (\bibinfo{year}{2001}),
  \urlprefix\url{http://link.aps.org/doi/10.1103/PhysRevLett.86.1195}.

\bibitem[{\citenamefont{Morizot et~al.}(2006)\citenamefont{Morizot, Colombe,
  Lorent, Perrin, and Garraway}}]{Morizot2006a}
\bibinfo{author}{\bibfnamefont{O.}~\bibnamefont{Morizot}},
  \bibinfo{author}{\bibfnamefont{Y.}~\bibnamefont{Colombe}},
  \bibinfo{author}{\bibfnamefont{V.}~\bibnamefont{Lorent}},
  \bibinfo{author}{\bibfnamefont{H.}~\bibnamefont{Perrin}}, \bibnamefont{and}
  \bibinfo{author}{\bibfnamefont{B.~M.} \bibnamefont{Garraway}},
  \bibinfo{journal}{Phys. Rev. A} \textbf{\bibinfo{volume}{74}},
  \bibinfo{pages}{023617} (\bibinfo{year}{2006}),
  \urlprefix\url{http://link.aps.org/doi/10.1103/PhysRevA.74.023617}.

\bibitem[{\citenamefont{Chakraborty et~al.}(2016)\citenamefont{Chakraborty,
  Mishra, Ram, Tiwari, and Rawat}}]{Chakraborty2016a}
\bibinfo{author}{\bibfnamefont{A.}~\bibnamefont{Chakraborty}},
  \bibinfo{author}{\bibfnamefont{S.~R.} \bibnamefont{Mishra}},
  \bibinfo{author}{\bibfnamefont{S.~P.} \bibnamefont{Ram}},
  \bibinfo{author}{\bibfnamefont{S.~K.} \bibnamefont{Tiwari}},
  \bibnamefont{and} \bibinfo{author}{\bibfnamefont{H.~S.} \bibnamefont{Rawat}},
  \bibinfo{journal}{Journal of Physics B: Atomic, Molecular and Optical
  Physics} \textbf{\bibinfo{volume}{49}}, \bibinfo{pages}{075304}
  (\bibinfo{year}{2016}),
  \urlprefix\url{http://stacks.iop.org/0953-4075/49/i=7/a=075304}.

\bibitem[{\citenamefont{Hess}(1986)}]{Hess1986a}
\bibinfo{author}{\bibfnamefont{H.~F.} \bibnamefont{Hess}},
  \bibinfo{journal}{Phys. Rev. B} \textbf{\bibinfo{volume}{34}},
  \bibinfo{pages}{3476} (\bibinfo{year}{1986}).

\bibitem[{\citenamefont{Ketterle and Druten}(1996)}]{Ketterle1996181}
\bibinfo{author}{\bibfnamefont{W.}~\bibnamefont{Ketterle}} \bibnamefont{and}
  \bibinfo{author}{\bibfnamefont{N.~V.} \bibnamefont{Druten}}
  (\bibinfo{publisher}{Academic Press}, \bibinfo{year}{1996}),
  vol.~\bibinfo{volume}{37} of \emph{\bibinfo{series}{Advances In Atomic,
  Molecular, and Optical Physics}}, pp. \bibinfo{pages}{181 -- 236},
  \urlprefix\url{http://www.sciencedirect.com/science/article/pii/S1049250X08601019}.

\bibitem[{\citenamefont{Anderson
  et~al.}(1995{\natexlab{b}})\citenamefont{Anderson, Ensher, Matthews, Wieman,
  and Cornell}}]{Anderson1996a}
\bibinfo{author}{\bibfnamefont{M.~H.} \bibnamefont{Anderson}},
  \bibinfo{author}{\bibfnamefont{J.~R.} \bibnamefont{Ensher}},
  \bibinfo{author}{\bibfnamefont{M.~R.} \bibnamefont{Matthews}},
  \bibinfo{author}{\bibfnamefont{C.~E.} \bibnamefont{Wieman}},
  \bibnamefont{and} \bibinfo{author}{\bibfnamefont{E.~A.}
  \bibnamefont{Cornell}}, in \emph{\bibinfo{booktitle}{Laser Spectroscopy,
  {XII} International Conference}}, edited by
  \bibinfo{editor}{\bibfnamefont{M.}~\bibnamefont{Inguscio}},
  \bibinfo{editor}{\bibfnamefont{M.}~\bibnamefont{Allegrini}},
  \bibnamefont{and} \bibinfo{editor}{\bibfnamefont{A.}~\bibnamefont{Sasso}}
  (\bibinfo{publisher}{World Scientific}, \bibinfo{year}{1995}{\natexlab{b}}),
  p.~\bibinfo{pages}{3}.

\bibitem[{\citenamefont{Barrett et~al.}(2001)\citenamefont{Barrett, Sauer, and
  Chapman}}]{Barrett2001a}
\bibinfo{author}{\bibfnamefont{M.~D.} \bibnamefont{Barrett}},
  \bibinfo{author}{\bibfnamefont{J.~A.} \bibnamefont{Sauer}}, \bibnamefont{and}
  \bibinfo{author}{\bibfnamefont{M.~S.} \bibnamefont{Chapman}},
  \bibinfo{journal}{Phys. Rev. Lett.} \textbf{\bibinfo{volume}{87}},
  \bibinfo{pages}{010404} (\bibinfo{year}{2001}),
  \urlprefix\url{http://link.aps.org/doi/10.1103/PhysRevLett.87.010404}.

\bibitem[{\citenamefont{Myatt et~al.}(1996)\citenamefont{Myatt, Newbury,
  Ghrist, Loutzenhiser, and Wieman}}]{Myatt1996a}
\bibinfo{author}{\bibfnamefont{C.~J.} \bibnamefont{Myatt}},
  \bibinfo{author}{\bibfnamefont{N.~R.} \bibnamefont{Newbury}},
  \bibinfo{author}{\bibfnamefont{R.~W.} \bibnamefont{Ghrist}},
  \bibinfo{author}{\bibfnamefont{S.}~\bibnamefont{Loutzenhiser}},
  \bibnamefont{and} \bibinfo{author}{\bibfnamefont{C.~E.}
  \bibnamefont{Wieman}}, \bibinfo{journal}{Opt. Lett.}
  \textbf{\bibinfo{volume}{21}}, \bibinfo{pages}{290} (\bibinfo{year}{1996}).

\bibitem[{\citenamefont{Mishra et~al.}(2008)\citenamefont{Mishra, Ram, Tiwari,
  and Mehendale}}]{SRMishra2008a}
\bibinfo{author}{\bibfnamefont{S.~R.} \bibnamefont{Mishra}},
  \bibinfo{author}{\bibfnamefont{S.~P.} \bibnamefont{Ram}},
  \bibinfo{author}{\bibfnamefont{S.~K.} \bibnamefont{Tiwari}},
  \bibnamefont{and} \bibinfo{author}{\bibfnamefont{S.~C.}
  \bibnamefont{Mehendale}}, \bibinfo{journal}{Phys. Rev. A}
  \textbf{\bibinfo{volume}{77}}, \bibinfo{pages}{065402}
  (\bibinfo{year}{2008}).

\bibitem[{\citenamefont{Esslinger et~al.}(1998)\citenamefont{Esslinger, Bloch,
  and H{\"a}nsch}}]{Esslinger1998a}
\bibinfo{author}{\bibfnamefont{T.}~\bibnamefont{Esslinger}},
  \bibinfo{author}{\bibfnamefont{I.}~\bibnamefont{Bloch}}, \bibnamefont{and}
  \bibinfo{author}{\bibfnamefont{T.~W.} \bibnamefont{H{\"a}nsch}},
  \bibinfo{journal}{Phys. Rev. A} \textbf{\bibinfo{volume}{58}},
  \bibinfo{pages}{R2664} (\bibinfo{year}{1998}).

\bibitem[{\citenamefont{Ram et~al.}(2010)\citenamefont{Ram, Tiwari, and
  Mishra}}]{Ram2010a}
\bibinfo{author}{\bibfnamefont{S.~P.} \bibnamefont{Ram}},
  \bibinfo{author}{\bibfnamefont{S.~K.} \bibnamefont{Tiwari}},
  \bibnamefont{and} \bibinfo{author}{\bibfnamefont{S.~R.}
  \bibnamefont{Mishra}}, \bibinfo{journal}{J. Korean Phys. Soc.}
  \textbf{\bibinfo{volume}{57}}, \bibinfo{pages}{1303} (\bibinfo{year}{2010}).

\bibitem[{\citenamefont{Ram et~al.}(2011)\citenamefont{Ram, Mishra, Tiwari, and
  Mehendale}}]{Ram2011a}
\bibinfo{author}{\bibfnamefont{S.~P.} \bibnamefont{Ram}},
  \bibinfo{author}{\bibfnamefont{S.~R.} \bibnamefont{Mishra}},
  \bibinfo{author}{\bibfnamefont{S.~K.} \bibnamefont{Tiwari}},
  \bibnamefont{and} \bibinfo{author}{\bibfnamefont{S.~C.}
  \bibnamefont{Mehendale}}, \bibinfo{journal}{Rev. Sci. Instrum.}
  \textbf{\bibinfo{volume}{82}}, \bibinfo{eid}{126108}
  (pages~\bibinfo{numpages}{3}) (\bibinfo{year}{2011}),
  \urlprefix\url{http://link.aip.org/link/?RSI/82/126108/1}.

\bibitem[{\citenamefont{Ram et~al.}(2013)\citenamefont{Ram, Tiwari, Mishra, and
  Rawat}}]{Ram2013a}
\bibinfo{author}{\bibfnamefont{S.~P.} \bibnamefont{Ram}},
  \bibinfo{author}{\bibfnamefont{S.~K.} \bibnamefont{Tiwari}},
  \bibinfo{author}{\bibfnamefont{S.~R.} \bibnamefont{Mishra}},
  \bibnamefont{and} \bibinfo{author}{\bibfnamefont{H.~S.} \bibnamefont{Rawat}},
  \bibinfo{journal}{Rev. Sci. Instrum.} \textbf{\bibinfo{volume}{84}},
  \bibinfo{eid}{073102} (pages~\bibinfo{numpages}{6}) (\bibinfo{year}{2013}),
  \urlprefix\url{http://link.aip.org/link/?RSI/84/073102/1}.

\bibitem[{\citenamefont{Ram}(2013)}]{Ram2013b}
\bibinfo{author}{\bibfnamefont{S.~P.} \bibnamefont{Ram}}, Ph.D. thesis,
  \bibinfo{school}{Homi Bhabha National Institute} (\bibinfo{year}{2013}).

\bibitem[{\citenamefont{Wang et~al.}(2008)\citenamefont{Wang, Wang, Yan, Geng,
  and Zhang}}]{JunminWang2008a}
\bibinfo{author}{\bibfnamefont{J.}~\bibnamefont{Wang}},
  \bibinfo{author}{\bibfnamefont{J.}~\bibnamefont{Wang}},
  \bibinfo{author}{\bibfnamefont{S.}~\bibnamefont{Yan}},
  \bibinfo{author}{\bibfnamefont{T.}~\bibnamefont{Geng}}, \bibnamefont{and}
  \bibinfo{author}{\bibfnamefont{T.}~\bibnamefont{Zhang}},
  \bibinfo{journal}{Rev. Sci. Instrum.} \textbf{\bibinfo{volume}{79}},
  \bibinfo{pages}{123116} (\bibinfo{year}{2008}).

\bibitem[{\citenamefont{Shu-Bin et~al.}(2006)\citenamefont{Shu-Bin, Tao,
  Tian-Cai, and Jun-Min}}]{Yan2006a}
\bibinfo{author}{\bibfnamefont{Y.}~\bibnamefont{Shu-Bin}},
  \bibinfo{author}{\bibfnamefont{G.}~\bibnamefont{Tao}},
  \bibinfo{author}{\bibfnamefont{Z.}~\bibnamefont{Tian-Cai}}, \bibnamefont{and}
  \bibinfo{author}{\bibfnamefont{W.}~\bibnamefont{Jun-Min}},
  \bibinfo{journal}{Chinese Phys.} \textbf{\bibinfo{volume}{15}},
  \bibinfo{pages}{1746} (\bibinfo{year}{2006}),
  \urlprefix\url{http://stacks.iop.org/1009-1963/15/i=8/a=019}.

\bibitem[{\citenamefont{Dimova et~al.}(2007)\citenamefont{Dimova, Morizot,
  Stern, Alzar, Fioretti, Lorent, Comparat, Perrin, and Pillet}}]{Dimova2007a}
\bibinfo{author}{\bibfnamefont{E.}~\bibnamefont{Dimova}},
  \bibinfo{author}{\bibfnamefont{O.}~\bibnamefont{Morizot}},
  \bibinfo{author}{\bibfnamefont{G.}~\bibnamefont{Stern}},
  \bibinfo{author}{\bibfnamefont{C.~G.} \bibnamefont{Alzar}},
  \bibinfo{author}{\bibfnamefont{A.}~\bibnamefont{Fioretti}},
  \bibinfo{author}{\bibfnamefont{V.}~\bibnamefont{Lorent}},
  \bibinfo{author}{\bibfnamefont{D.}~\bibnamefont{Comparat}},
  \bibinfo{author}{\bibfnamefont{H.}~\bibnamefont{Perrin}}, \bibnamefont{and}
  \bibinfo{author}{\bibfnamefont{P.}~\bibnamefont{Pillet}},
  \bibinfo{journal}{Eur. Phys. J. D} \textbf{\bibinfo{volume}{42}},
  \bibinfo{pages}{299} (\bibinfo{year}{2007}).

\bibitem[{\citenamefont{Swanson et~al.}(1998)\citenamefont{Swanson, Asgeirsson,
  Behr, Gorelov, and Melconian}}]{Swanson1998a}
\bibinfo{author}{\bibfnamefont{T.~B.} \bibnamefont{Swanson}},
  \bibinfo{author}{\bibfnamefont{D.}~\bibnamefont{Asgeirsson}},
  \bibinfo{author}{\bibfnamefont{J.~A.} \bibnamefont{Behr}},
  \bibinfo{author}{\bibfnamefont{A.}~\bibnamefont{Gorelov}}, \bibnamefont{and}
  \bibinfo{author}{\bibfnamefont{D.}~\bibnamefont{Melconian}},
  \bibinfo{journal}{J. Opt. Soc. Am. B} \textbf{\bibinfo{volume}{15}},
  \bibinfo{pages}{2641} (\bibinfo{year}{1998}).

\bibitem[{\citenamefont{Schaff et~al.}(2011)\citenamefont{Schaff, Capuzzi,
  Labeyrie, and Vignolo}}]{Schaff2011a}
\bibinfo{author}{\bibfnamefont{J.-F.} \bibnamefont{Schaff}},
  \bibinfo{author}{\bibfnamefont{P.}~\bibnamefont{Capuzzi}},
  \bibinfo{author}{\bibfnamefont{G.}~\bibnamefont{Labeyrie}}, \bibnamefont{and}
  \bibinfo{author}{\bibfnamefont{P.}~\bibnamefont{Vignolo}},
  \bibinfo{journal}{New Journal of Physics} \textbf{\bibinfo{volume}{13}},
  \bibinfo{pages}{113017} (\bibinfo{year}{2011}),
  \urlprefix\url{http://stacks.iop.org/1367-2630/13/i=11/a=113017}.

\bibitem[{\citenamefont{Meyrath}(2005)}]{Meyrath2005a}
\bibinfo{author}{\bibfnamefont{T.~P.} \bibnamefont{Meyrath}}, Ph.D. thesis,
  \bibinfo{school}{University of {T}exas at {A}ustin} (\bibinfo{year}{2005}).

\bibitem[{\citenamefont{Yoon}(2009)}]{Yoon2009a}
\bibinfo{author}{\bibfnamefont{M.~S.} \bibnamefont{Yoon}}, Ph.D. thesis,
  \bibinfo{school}{University of Oxford} (\bibinfo{year}{2009}).

\bibitem[{\citenamefont{Petrich et~al.}(1995)\citenamefont{Petrich, Anderson,
  Ensher, and Cornell}}]{Petrich1995a}
\bibinfo{author}{\bibfnamefont{W.}~\bibnamefont{Petrich}},
  \bibinfo{author}{\bibfnamefont{M.~H.} \bibnamefont{Anderson}},
  \bibinfo{author}{\bibfnamefont{J.~R.} \bibnamefont{Ensher}},
  \bibnamefont{and} \bibinfo{author}{\bibfnamefont{E.~A.}
  \bibnamefont{Cornell}}, \bibinfo{journal}{Phys. Rev. Lett.}
  \textbf{\bibinfo{volume}{74}}, \bibinfo{pages}{3352} (\bibinfo{year}{1995}),
  \urlprefix\url{http://link.aps.org/doi/10.1103/PhysRevLett.74.3352}.

\bibitem[{\citenamefont{Pritchard}(1983)}]{Pritchard1983a}
\bibinfo{author}{\bibfnamefont{D.~E.} \bibnamefont{Pritchard}},
  \bibinfo{journal}{Phys. Rev. Lett.} \textbf{\bibinfo{volume}{51}},
  \bibinfo{pages}{1336} (\bibinfo{year}{1983}).

\bibitem[{\citenamefont{Lu and van Wijngaarden}(2004)}]{Lu2004a}
\bibinfo{author}{\bibfnamefont{B.}~\bibnamefont{Lu}} \bibnamefont{and}
  \bibinfo{author}{\bibfnamefont{W.~A.} \bibnamefont{van Wijngaarden}},
  \bibinfo{journal}{Can. J. Phys.} \textbf{\bibinfo{volume}{82}},
  \bibinfo{pages}{81} (\bibinfo{year}{2004}).

\bibitem[{\citenamefont{Ketterle et~al.}(1999)\citenamefont{Ketterle, Durfee,
  and Stamper-Kurn}}]{Ketterle1999a}
\bibinfo{author}{\bibfnamefont{W.}~\bibnamefont{Ketterle}},
  \bibinfo{author}{\bibfnamefont{D.~S.} \bibnamefont{Durfee}},
  \bibnamefont{and} \bibinfo{author}{\bibfnamefont{D.~M.}
  \bibnamefont{Stamper-Kurn}}, in \emph{\bibinfo{booktitle}{Proceedings of the
  International School of Physics - Enrico Fermi}}, edited by
  \bibinfo{editor}{\bibfnamefont{M.}~\bibnamefont{Inguscio}},
  \bibinfo{editor}{\bibfnamefont{S.}~\bibnamefont{Stringari}},
  \bibnamefont{and} \bibinfo{editor}{\bibfnamefont{C.~E.} \bibnamefont{Wieman}}
  (\bibinfo{publisher}{IOS Press}, \bibinfo{year}{1999}),
  p.~\bibinfo{pages}{67}.

\bibitem[{\citenamefont{Stamper-Kurn et~al.}(1998)\citenamefont{Stamper-Kurn,
  Miesner, Inouye, Andrews, and Ketterle}}]{Stamper-Kurn1998a}
\bibinfo{author}{\bibfnamefont{D.~M.} \bibnamefont{Stamper-Kurn}},
  \bibinfo{author}{\bibfnamefont{H.-J.} \bibnamefont{Miesner}},
  \bibinfo{author}{\bibfnamefont{S.}~\bibnamefont{Inouye}},
  \bibinfo{author}{\bibfnamefont{M.~R.} \bibnamefont{Andrews}},
  \bibnamefont{and} \bibinfo{author}{\bibfnamefont{W.}~\bibnamefont{Ketterle}},
  \bibinfo{journal}{Phys. Rev. Lett.} \textbf{\bibinfo{volume}{81}},
  \bibinfo{pages}{500} (\bibinfo{year}{1998}),
  \urlprefix\url{http://link.aps.org/doi/10.1103/PhysRevLett.81.500}.

\bibitem[{\citenamefont{Bradley et~al.}(1995)\citenamefont{Bradley, Sackett,
  Tollett, and Hulet}}]{Bradley1995a}
\bibinfo{author}{\bibfnamefont{C.~C.} \bibnamefont{Bradley}},
  \bibinfo{author}{\bibfnamefont{C.~A.} \bibnamefont{Sackett}},
  \bibinfo{author}{\bibfnamefont{J.~J.} \bibnamefont{Tollett}},
  \bibnamefont{and} \bibinfo{author}{\bibfnamefont{R.~G.} \bibnamefont{Hulet}},
  \bibinfo{journal}{Phys. Rev. Lett.} \textbf{\bibinfo{volume}{75}},
  \bibinfo{pages}{1687} (\bibinfo{year}{1995}), \bibinfo{note}{{\it{ibid.}
  \bf{79}}, 1170 (1997)}.

\bibitem[{\citenamefont{Yu-Zhu et~al.}(2003)\citenamefont{Yu-Zhu, Shu-Yu, Quan,
  Shan-Yu, and Hai-Xiang}}]{Yu-Zhu2003a}
\bibinfo{author}{\bibfnamefont{W.}~\bibnamefont{Yu-Zhu}},
  \bibinfo{author}{\bibfnamefont{Z.}~\bibnamefont{Shu-Yu}},
  \bibinfo{author}{\bibfnamefont{L.}~\bibnamefont{Quan}},
  \bibinfo{author}{\bibfnamefont{Z.}~\bibnamefont{Shan-Yu}}, \bibnamefont{and}
  \bibinfo{author}{\bibfnamefont{F.}~\bibnamefont{Hai-Xiang}},
  \bibinfo{journal}{Chinese Physics Letters} \textbf{\bibinfo{volume}{20}},
  \bibinfo{pages}{799} (\bibinfo{year}{2003}),
  \urlprefix\url{http://stacks.iop.org/0256-307X/20/i=6/a=306}.

\bibitem[{\citenamefont{Zhang et~al.}(2005)\citenamefont{Zhang, Xu, and
  You}}]{Zhang2005a}
\bibinfo{author}{\bibfnamefont{W.}~\bibnamefont{Zhang}},
  \bibinfo{author}{\bibfnamefont{Z.}~\bibnamefont{Xu}}, \bibnamefont{and}
  \bibinfo{author}{\bibfnamefont{L.}~\bibnamefont{You}},
  \bibinfo{journal}{Phys. Rev. A} \textbf{\bibinfo{volume}{72}},
  \bibinfo{pages}{053627} (\bibinfo{year}{2005}),
  \urlprefix\url{http://link.aps.org/doi/10.1103/PhysRevA.72.053627}.

\bibitem[{\citenamefont{Durfee and Ketterle}(1998)}]{Durfee1998a}
\bibinfo{author}{\bibfnamefont{D.~S.} \bibnamefont{Durfee}} \bibnamefont{and}
  \bibinfo{author}{\bibfnamefont{W.}~\bibnamefont{Ketterle}},
  \bibinfo{journal}{Opt. Express} \textbf{\bibinfo{volume}{2}},
  \bibinfo{pages}{299} (\bibinfo{year}{1998}),
  \urlprefix\url{http://www.opticsexpress.org/abstract.cfm?URI=oe-2-8-299}.

\bibitem[{\citenamefont{Naraschewski and
  Stamper-Kurn}(1998)}]{Naraschewski1998a}
\bibinfo{author}{\bibfnamefont{M.}~\bibnamefont{Naraschewski}}
  \bibnamefont{and} \bibinfo{author}{\bibfnamefont{D.~M.}
  \bibnamefont{Stamper-Kurn}}, \bibinfo{journal}{Phys. Rev. A}
  \textbf{\bibinfo{volume}{58}}, \bibinfo{pages}{2423} (\bibinfo{year}{1998}),
  \urlprefix\url{http://link.aps.org/doi/10.1103/PhysRevA.58.2423}.

\end{thebibliography}

\end{document}